\documentclass[useAMS]{gGAF2e}

\begin{document}
\doi{10.1080/03091920xxxxxxxxx}
 \issn{1029-0419} \issnp{0309-1929} %\jvol{00} \jnum{00} \jyear{2006} \jmonth{February}

%\markboth{\LaTeXe\ guide for authors}{\LaTeXe\ guide for authors}   % short title
\markboth{Magnetohydrodynamics of the elliptical instability}
     {Magnetohydrodynamics of the elliptical instability}

\title{{\textit{Magnetohydrodynamic simulations of the elliptical instability in triaxial ellipsoids}}}

\author{D. C\'ebron,
\thanks{$^\ast$
Corresponding author. Email: cebron@irphe.univ-mrs.fr
\vspace{6pt}
\newline\centerline{\tiny{{\em Geophysical and Astrophysical Fluid Dynamics}}}
\newline\centerline{\tiny{  ISSN 0309-1929 print/ ISSN 1029-0419 online \textcopyright 2006 Taylor \& Francis Ltd}}
\newline\centerline{\tiny{ http://www.tandf.co.uk/journals}}
\newline \centerline{\tiny{DOI:10.1080/03091920xxxxxxxxx}}}
M. Le Bars, P. Maubert and P. Le Gal \\
\vspace{6pt} Aix-Marseille Univ., IRPHE (UMR 6594), 13384, Marseille cedex 13, France.
\received{received xxx}}

\maketitle

\begin{abstract}
The elliptical instability can take place in planetary cores
and stars elliptically deformed by gravitational effects, where it
generates large-scale three-dimensional flows assumed to be dynamo
capable. In this work, we present the first magneto-hydrodynamic
numerical simulations of such flows, using a finite-element method.
We first validate our numerical approach by comparison with
kinematic and dynamic dynamos benchmarks of the literature. We then
systematically study the magnetic field induced by various modes of
the elliptical instability from an imposed external field in a
triaxial ellipsoidal geometry, relevant in a geo- and astrophysical
context. Finally, in tidal induction cases, the external magnetic field is suddenly shut down and the decay rates of the magnetic field are systematically reported.
\end{abstract}

\section{Introduction} \label{sec:intro}

Many celestial bodies (planets, stars, galaxies...) possess their
own magnetic field, either by induction from an external field, or by
a natural dynamo mechanism. Up to now, only two kinds of natural
forcing have been identified as dynamo-capable in celestial bodies: (i) thermo-solutal convection \cite[][]{Glatzmaier_1995}, which is the
standard mechanism generally applied to all planetary configurations
even if it is not proved to be always relevant; and (ii) precession \cite[][]{Tilgner_2005}, a purely mechanical forcing that
may drive dynamos in some planets and moons \cite[][]{malkus1968},
despite a well-known controversy on its energetic budget (see \citealt{Rochester_1975,Loper_1975} for critisms of this hypothesis, and \citealt{Kerswell_1996} for its rehabilitation). On Earth today, the magnetic field is very likely generated by thermochemical
convective motions within the electrically conducting liquid core, driven by the
solidification of its inner core. However, the origin of the
magnetic field in the Early Earth, in the Moon, in Ganymede or in
Mars is more uncertain, and leads to the consideration of
alternative dynamo mechanisms \cite[][]{Jones_2003,Jones_2011}. The recent discovery
of fast magnetic reversals on the extra-solar star Tau-boo \cite[][]{Donati,Fares}, which may be related to strong tidal effects due to the
presence of a massive close companion \cite[][]{Fares}, also requires
to re-evaluate classical models of convective dynamos. Indeed, even
when the dynamo is of a convective origin, the role of other driving
mechanisms can be very important in the organization of fluid
motions.

In addition to convection and precession, two other mechanisms have also been proposed to be of fundamental importance to
drive cores flows, and consequently to influence planetary magnetic fields: libration and tides. As recently shown in \cite{Cebron_2011},
both these forcings are indeed capable of extracting huge amount
of rotational energy to create complex three-dimensional motions
through the excitation of a so-called elliptical instability, also
called tidal instability in the astrophysical context. The
elliptical instability is a generic instability that affects any
rotating fluid whose streamlines are elliptically deformed \cite[see
for instance the review by][]{Kerswell_2002}. A fully
three-dimensional turbulent flow is excited in the bulk as soon as
(i) the ratio between the ellipticity $\beta$ of the streamlines and
the square root of the Ekman number $E$ is larger than a critical
value of order one and (ii) a difference in angular velocity exists
between the mean rotation of the fluid and the elliptical
distortion. In a planetary context, the ellipticity of streamlines
is related to the tidal deformation of the planetary layers. A differential rotation is generically present between the core fluid
and the dynamic tides in non-synchronized systems; it also appears
between the static bulge and the core fluid because of librations in
synchronized ones. The elliptical instability is then refereed to tide driven elliptical instability (TDEI) and libration driven
elliptical instability (LDEI), respectively.

So far, magneto-hydrodynamic (MHD) simulations of stellar or
planetary flows have been performed in spherical, and recently
spheroidal (i.e. axisymetric around the rotation axis), geometries, which facilitate and accelerate the
computations but also prevent the growth of any elliptical instability. Because of
the small amplitudes of tidal bulges, this approximation could be
thought to be correct. But since the elliptical instability comes
from a parametric resonance, even an infinitesimal deformation can
lead to first order modifications of the flow \cite[][]{Lacaze_2004,Cebron_2010a}. To study its MHD
consequences, we have thus developed the first numerical MHD
simulations in a triaxial ellipsoidal geometry, which are presented
below.

The paper is organized as follows. In section \ref{sec:num}, the
numerical method used to solve MHD flows in a non-axisymmetric
geometry is described. In section \ref{sec:valid}, validations of
our method are presented, considering kinematic dynamos with
different mesh elements and boundary conditions, but also a
thermally driven dynamic dynamo following the standard benchmark
proposed by \cite{Christensen}. In section \ref{sec:MHD_elliptic}, the code is finally used to study induction processes by the elliptical instability. These simulations are then used to systematically study the decay rates of the magnetic field when the external magnetic field is suddenly shut down.

\section{Numerical model} \label{sec:num}

\subsection{Local methods in MHD simulations}

Since the pioneering work of \cite{Glatzmaier_1995}, numerical
simulations have more and more deeply studied how a convective dynamo process
can generate a magnetic field in a rotating shell. Many comparisons
with the observational data, mainly obtained on Earth but also on other planets of
the solar system, have been done to confirm the relevance of these
simulations \cite[e.g.][for a review]{Dormy_2000}. Some key features like the dipole
dominance, the westward drift and the occasional reversals of the
magnetic field, are recovered by various codes. However, due to computational costs, the dimensionless parameters used in numerical
simulations are very different from the realistic values of
planetary cores \cite[][]{Busse_2002}. To get as close as possible
to the real parameter values, the numerical codes are optimized and
massively parallel. Usually, numerical simulations of
magneto-hydrodynamic flows in planetary cores benefit from their
spherical geometry to use fast and precise spectral methods.
However, since global communication is required
\cite[e.g.][]{Clune_1999}, such methods are difficult to
parallelize. Following the precursory work of \cite{Kageyama_1997},
some studies have been performed using local methods, more easily
adapted to parallel architectures, and thus more suitable for
massively parallel computations: see for instance the works of
\cite{Chan_2001,Matsui_2004,Matsui_2005} using finite-element
methods; the works of \cite{Hejda_2003,Hejda_2004,Harder_2005} using
finite-volume methods; and the work of \cite{Fournier_2004,Fournier_2005} using
spectral elements. Besides, even if all these previous studies have
been performed in spheres, local methods also have the great
advantage of providing robust and accurate solutions for arbitrary
geometries, which is of direct interest for our study of flows
driven in triaxial ellipsoidal geometries.

\subsection{MHD equations} \label{sec:pb}

We consider a finite volume of conducting fluid of kinematic
viscosity $\nu$, density $\rho$, magnetic diffusivity $\nu_m$ and
electrical conductivity $\gamma$, rotating with a typical rate
$\Omega$ and enclosed in a rigid container of typical size $R$.
Using $R$ as a lengthscale, $\Omega^{-1}$ as a time scale, and
$\Omega\ R \sqrt{\rho \mu_0}$ as a magnetic field scale, where
$\mu_0$ is the vacuum magnetic permeability, the
magnetohydrodynamics equations in the non-relativistic limit
(equivalently, considering only timescales greater than the
relaxation time of the charge carriers) write
\begin{eqnarray}
\frac{\partial \mathbf{u}}{\partial t}+ \mathbf{u} \cdot
\mathbf{\nabla} \mathbf{u} &=& -\mathbf{\nabla} p + E
\boldsymbol{\bigtriangleup} \mathbf{u} - 2\ \boldsymbol{\Omega}\times \boldsymbol{u}+  (\mathbf{\nabla \times B}) \times \mathbf{B_{tot}}, \label{U1}\\
\mathbf{\nabla}  \cdot \mathbf{u} &=& 0, \label{U2}\\
\frac{\partial \mathbf{B}}{\partial t} &=&  \mathbf{\nabla \times (u
\times B_{tot})}  + \frac{1}{Rm}
\boldsymbol{\bigtriangleup} \mathbf{B} \label{induc_B}, \\
\mathbf{\nabla}  \cdot \mathbf{B} &=&0.\label{div_free}
\end{eqnarray}
where $(\mathbf{u}, p, \mathbf{B})$ are respectively the velocity,
pressure and magnetic fields, and
$\mathbf{B_{tot}}=\mathbf{B}+\mathbf{B_0}$ is the total magnetic
field accounting for a possible constant and uniform
$\mathbf{B_0}$, representing the imposed external magnetic field
in induction problems. The right-hand side term $(\mathbf{\nabla \times B}) \times \mathbf{B_{tot}}$ in equation (\ref{U1}) corresponds to the so-called Laplace (or Lorentz) force. Dimensionless numbers are the Ekman number $E=\nu/(\Omega\ R^2)$, and the magnetic Reynolds number $Rm=Pm/E$,
where $Pm=\nu/\nu_m$ is the magnetic Prandtl number. In this work,
the no-slip boundary condition is systematically used for the fluid.
Note that a Coriolis force $- 2\ \boldsymbol{\Omega} \times
\boldsymbol{u}$, where $\boldsymbol{\Omega}$ is the rotation vector
of the working frame of reference, is introduced here for generality
and will be used in section \ref{valid4}. Once the magnetic field is
solved, the Maxwell's system of equations allows us to deduce the
current density $\mathbf{j}=\mathbf{\nabla \times B}/\mu_0 $, the
electric potential $\bigtriangleup V_e= \mathbf{\nabla \cdot (u
\times B)}$ and the charges distribution $\rho_e=- \epsilon
\mathbf{\nabla \cdot (u \times B)}$, where $\epsilon$ is the
electric permittivity of the fluid.

Usually, numerical simulations of magneto-hydrodynamic flows in
planetary cores benefit from their spherical geometry to use fast
and precise spectral methods. In our case however, we do not
consider any simple symmetry. Our computations are thus performed
with a standard finite-element method, widely used in engineering
studies, which allows to deal with complex geometries and to simply
impose the fluid boundary conditions. However, solving the magnetic
field with local methods gives rise to some difficulties that we
have to cope with. In the finite-element community, the MHD
simulations are usually done with a formulation in terms of magnetic
potential vector defined by $\mathbf{B}= \mathbf{\nabla \times A}$
\cite[see e.g.][]{Matsui_2005}, which ensures that the field remains
divergence free at any time. Equations
(\ref{induc_B}-\ref{div_free}) are thus replaced by
\begin{eqnarray}
\frac{\partial \mathbf{A}}{\partial t} &=& (\mathbf{u} \times
\mathbf{B_{tot}}) + \frac{1}{Rm}
\boldsymbol{\bigtriangleup} \mathbf{A},    \label{induc_A} \\
\mathbf{B_{tot}} &=& \mathbf{\nabla \times A}+ \mathbf{B_0}=
\mathbf{B}+ \mathbf{B_0}. \label{induc_A2}
\end{eqnarray}
Naturally, the two MHD equations (\ref{induc_B}-\ref{div_free}) can
be recovered from equations (\ref{induc_A}-\ref{induc_A2}). The
absence of gauge, usually introduced in potential vector formulation
\cite[e.g.][]{Matsui_2005}, prevents us from dealing with purely
insulating domain. Indeed, in such domains, the electric field is
the Lagrange multiplier associated with the constraint
$\boldsymbol{\nabla \times B} = \boldsymbol{0}$
\cite[e.g.][]{Guermond_2007}. This imposes to use a supplementary
variable $\phi$, which is a Lagrange multiplier to ensure the gauge
$ \mathbf{\nabla}  \cdot \mathbf{A} =0$. The variable $\phi$ is equivalent to the pressure $p$ for the velocity field. Consequently, since we do not impose any gauge, we cannot use perfectly insulating materials in the following. Thus, we model them by very weakly conducting domains compared to the metallic ones.

\subsection{Boundary conditions} \label{sec:BC}

The non-local nature of the magnetic boundary conditions is a
long-standing issue in dynamo modeling. Usually with spectral
methods, the matching of the magnetic induction at the boundaries
with an outer potential field is easily tractable. However,
using local methods, a main issue remains regarding the conflict
between local discretization and the global form of the magnetic
boundary conditions. As reviewed in \cite{Iskakov_2004}, different
solutions have been proposed in the literature to cope with this
problem. The so-called quasi-vacuum condition $\mathbf{n \times
B=0}$ ($\mathbf{n}$ being the local normal vector), which is a local condition representing for instance an
outside domain made of a perfect 'magnetic conductor' ($\mu \rightarrow \infty$), has been used
by \cite{Kageyama_1997} and \cite{Harder_2005}. The immersion
of the bounded domain into a large domain where the magnetic problem
is also solved has been used by \cite{Chan_2001} and improved by \cite{Matsui_2004} in using the
potential vector. The combination of the local method with an
integral boundary elements method (BEM) has been introduced by
\cite{Iskakov_2004} by coupling a finite volume method with the
BEM. Here, following \cite{Matsui_2004}, we choose solutions (i) or (ii),
depending on the considered problem. When the solution (ii) is used,
we impose on a distant external spherical boundary the magnetic
condition $\mathbf{A \times n =0}$, which corresponds to $\mathbf{B
\cdot n} =0$. 

\subsection{Numerical method}

To solve the complete MHD problem, we use the commercial software
COMSOL Multiphysics\textsuperscript{\circledR}. For the fluid
variables, the mesh element type is the standard Lagrange element
$P1-P2$, which is linear for the pressure field and quadratic for
the velocity field. Note that higher order elements such as $P2-P3$ would have a better convergence rate with the number of mesh elements but would impose a significant supplementary computational cost. For a given computational cost, the use of $P1-P2$ elements allows to use a finer mesh. 

Lagrange elements are nodal elements, well
adapted to solve for the velocity field. However, as reminded by \cite{Hesthaven_2004}, the use of this
kind of elements in a straightforward nodal continuous Galerkin
finite-element method is known to lead to the appearance of
spurious, non-physical solutions \cite[e.g.][for a review]{Sun_1995,Jiang_1996}.
Their origin has several interpretations such as a poor
representation of the large null space of the involved operator
\cite[e.g.][]{Bossavit_1988} or the generation of solutions that
violate the divergence conditions, which are typically not imposed
directly \cite[][]{Paulsen_1991}. Another difficulty is that the
singular component of the solution may be not computed if the
interface between a conductive medium and a non-conductive medium is
not smooth \cite[see the lemma of ][]{Costabel_1991}. Finally,
difficulties appear on the coupling of fields across such an
interface. A first way to overcome these difficulties is to use
specific method such as interior penalty discontinuous Galerkin
methods \cite[e.g.][for MHD applications]{Guermond_2007}. A second
way to solve this problem is to construct elements adapted to the operator. In a
pionnering work, \cite{Bossavit_1988,Bossavit_1990} shows that the
use of special curl-conforming elements
\cite[][]{Nedelec_1980,Nedelec_1986} allows to overcome the problem
of spurious modes. Finite-element methods based on such curl-conforming elements, also
called N\'ed\'elec's edge (or vector) elements, constitute now the
dominating approach for solving geometrically complex problems
\cite[e.g.][]{Jin_1993, Volakis_1998}. An important advantage of
edge elements is that they ensure the continuity of tangential field
components across an interface between different media, while
leaving the normal field components free to jump across such
interfaces, which is a typical property of  electromagnetic problems
\cite[see also][for details]{Monk_2003}. This also implies that the
curl of the vector field is an integrable function, so these
elements are suitable for equations using the curl of the vector
field, such as the potential vector. 

In this work, the mesh element
type employed for the magnetic potential vector is thus the
N\'ed\'elec edge element, either linear or quadratic depending on
the considered problem. The number of degrees of freedom (DoF) used
in most simulations of this work ranges between $5\cdot10^4$
DoF for kinematic dynamos and $8\cdot10^5$ DoF for full MHD problems
with a magnetic Reynolds number about $10^3$. We use the so-called Implicit Differential-Algebraic solver (IDA solver), based on backward differencing formulas (see Hindmarsh et al. 2005 for details on the IDA solver). The integration method in IDA is variable-order (and variable-coefficient BDF), the order ranging between 1 and 5. At each time step the system is solved with the sparse direct linear solver PARDISO (www.pardiso-project.org). Up to know, the
commercial software COMSOL Multiphysics\textsuperscript{\circledR}
was not parallelized, and all computations were performed on a
single workstation with 96 Go RAM, and two processors Intel\textsuperscript{\circledR} Xeon\textsuperscript{\circledR} E5520 (2.26 GHz, 8MB Cache). Note that each numerical simulation presented in this work typically requires 64 Go RAM and was performed on a single processor, which leads to typical CPU times of half a day for kinematic dynamos and induction calculations and CPU time of weeks when the full dynamo problem is solved. Note also that the latest version of COMSOL Multiphysics\textsuperscript{\circledR} delivered in the summer 2011 should allow parallelized calculations and we hope to access to a significantly increased numerical power very
soon. In addition to their purely scientific interest, results
presented in this paper should thus be considered as a first step
towards solving MHD numerical problems with this commercial
software.

\section{Validation of the model} \label{sec:valid}

\subsection{Ponomarenko-like dynamo problem}\label{valid2}

\begin{figure}
  \begin{center}
    \begin{tabular}{cc}
      \setlength{\epsfysize}{5.5cm}
      \subfigure[]{\epsfbox{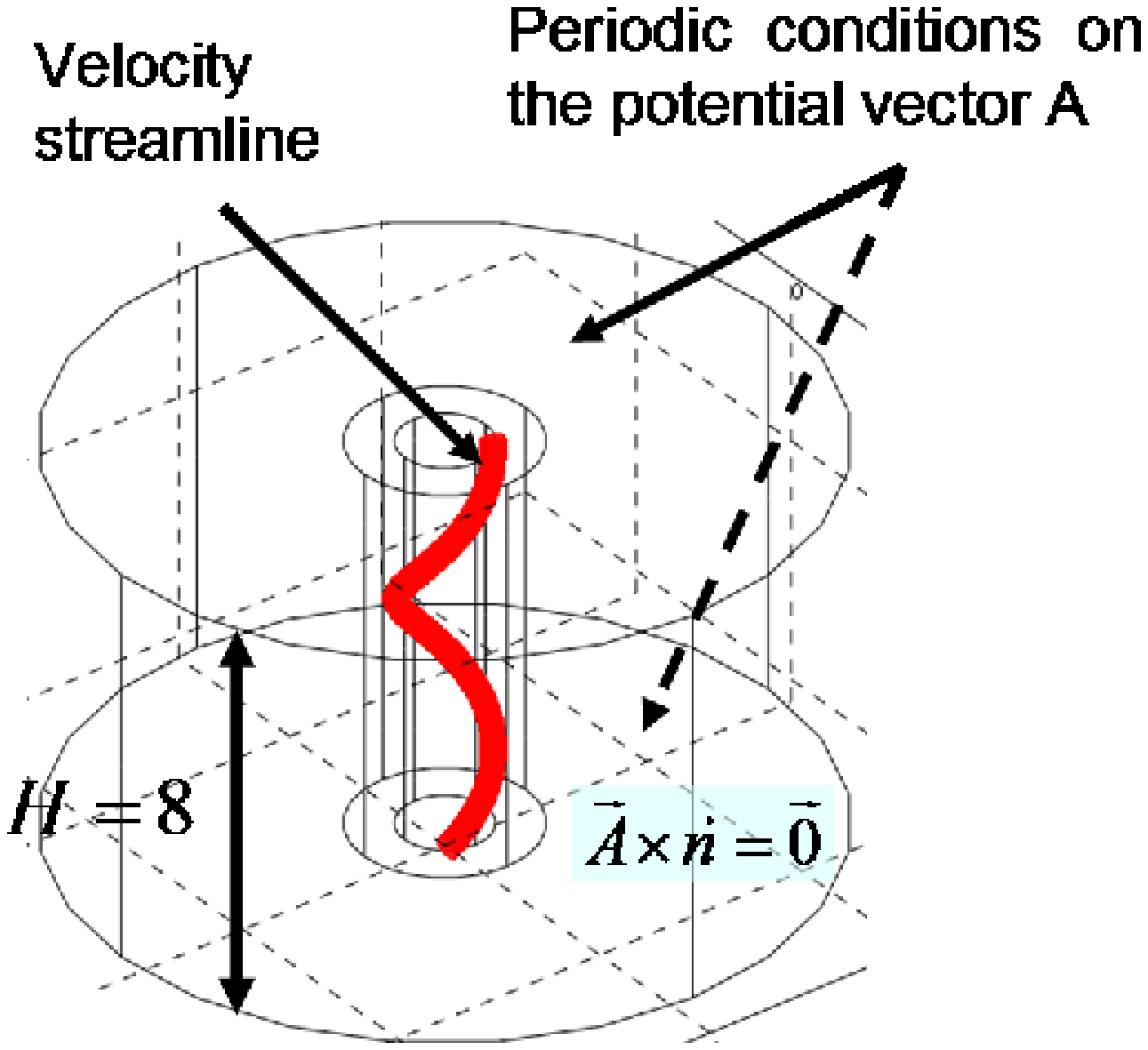}} &
      \setlength{\epsfysize}{5.1cm}
      \subfigure[]{\epsfbox{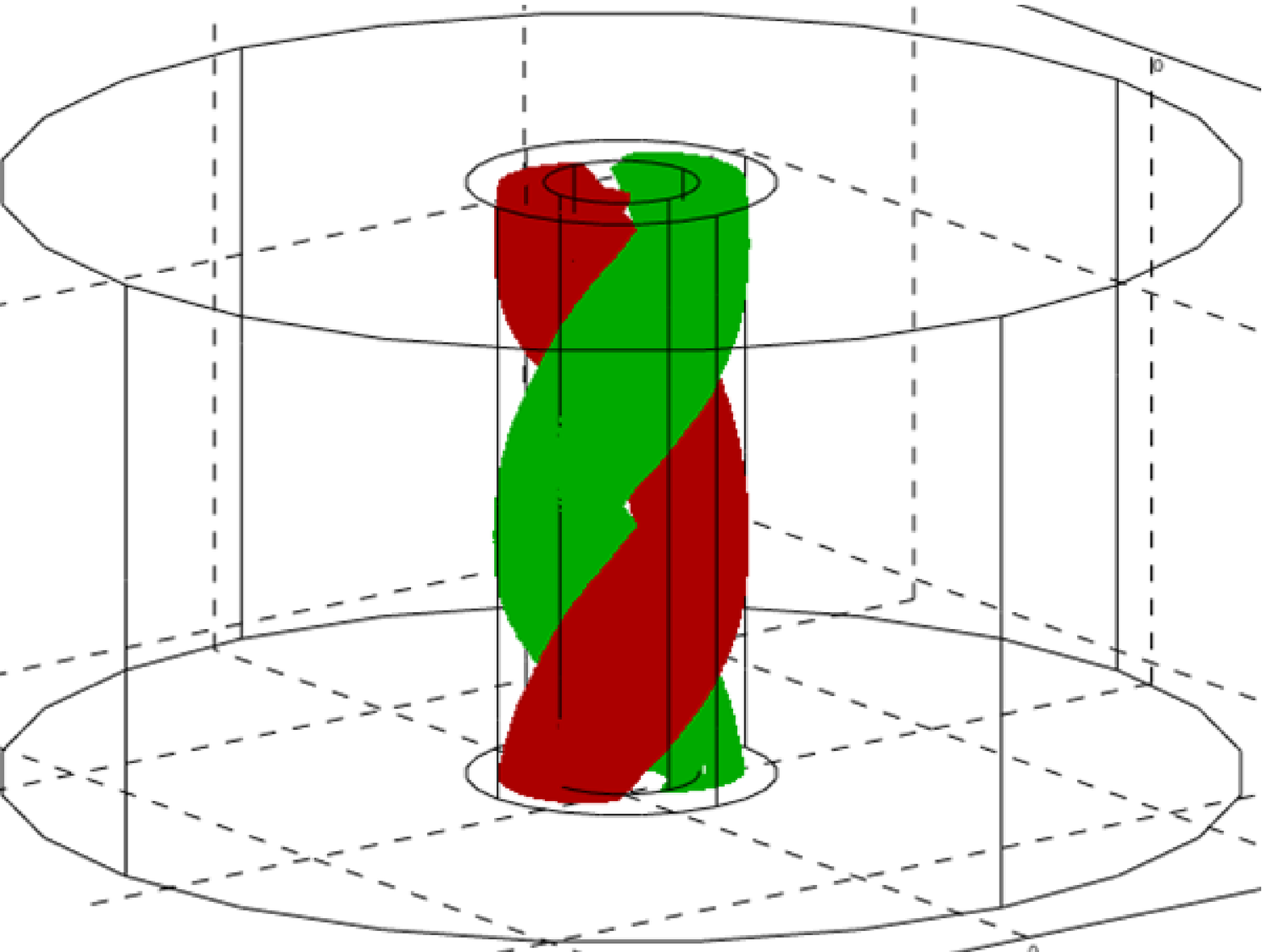}}
    \end{tabular}
    \caption{(a) Sketch of the Ponomarenko-like problem solved in the numerical simulations.
    In the region $r \leq 1$, the helical flow (\ref{pono_flow}) is imposed; in the region $r \in [1; 2]$,
    the fluid is supposed at rest; and the region $r \in [2; 8]$ is assumed to be insulating. (b) Iso-surfaces of the axial component of the magnetic field with $25\ \%$ of the maximum (red)
    and $25\ \%$ of the minimum (green) determined with our code at $R_m=20$.}.
    \label{cebronfig1}
  \end{center}
\end{figure}

We consider in this section a Ponomarenko-like configuration, which
is a well-known kinematic dynamo. In his original formulation,
\cite{Pono_1973} considered the flow of an electrically conductive
fluid within a cylinder of radius $R$, immersed into an infinite
conductive medium at rest. The flow is a solid body screw motion,
defined in cylindrical coordinates $(r,\theta,z)$ by
\begin{eqnarray} \mathbf{u}=
\left [
   \begin{array}{ccc}
      u_r\\
      u_{\theta} \\
      u_z\\
   \end{array}
   \right ]
=
\left [
   \begin{array}{ccc}
      0\\
      \Omega\ r \\
      R_b\ \Omega R\\
   \end{array}
   \right ], \label{pono_flow}
\end{eqnarray}
where $R_b$ is the ratio between the axial velocity and the rotation
velocity at the boundary $r=R$, i.e. the pitch of the spiral. We
choose $R$ and $\Omega^{-1}$ as the lengthscale and the timescale,
respectively. With this imposed flow, the kinematic dynamo problem is
analytically tractable and the critical eigenmode associated to the
smallest magnetic Reynolds number $R_m^c=17.73$ corresponds to
$R_b=1.3$, $k=-0.39$ and $m=1$, where $k$ and $m$ are respectively
the axial and azimuthal wavenumbers of the solution.

As sketched in figure \ref{cebronfig1}a, we consider the slightly
different case of an helical flow immersed into a stagnant
conductive medium with the same electrical conductivity in the
region $r \in [1; 2]$, and an insulating region $2 \leq r \leq 8$.
We use $R_b=1$ and a height $H=8$, similar to
\cite{Kaiser_1999,Laguerre_2006}. The boundary conditions are
$\mathbf{A \times n=0}$ on the outer sidewall, and we impose
periodicity of the potential vector on the top/bottom of the
cylinder. As explained in section \ref{sec:pb}, the insulating domain
$2 \leq r \leq 8$ is replaced in our code by a domain $10^{-8}$
times less conductive than the fluid domain. Concerning the
validation of the numerical code, the interest of this problem is
threefold: comparing our results with thoses of the literature, we
can estimate (i) the influence of the non-zero outer conductivity,
(ii) the capacity of the code to solve discontinuities of the flow
and of the electrical conductivity, and (iii) the relevance of our
boundary conditions.

\begin{figure}                   
  \begin{center}
    \begin{tabular}{ccc}
      \setlength{\epsfysize}{6.5cm}
      \subfigure[]{\epsfbox{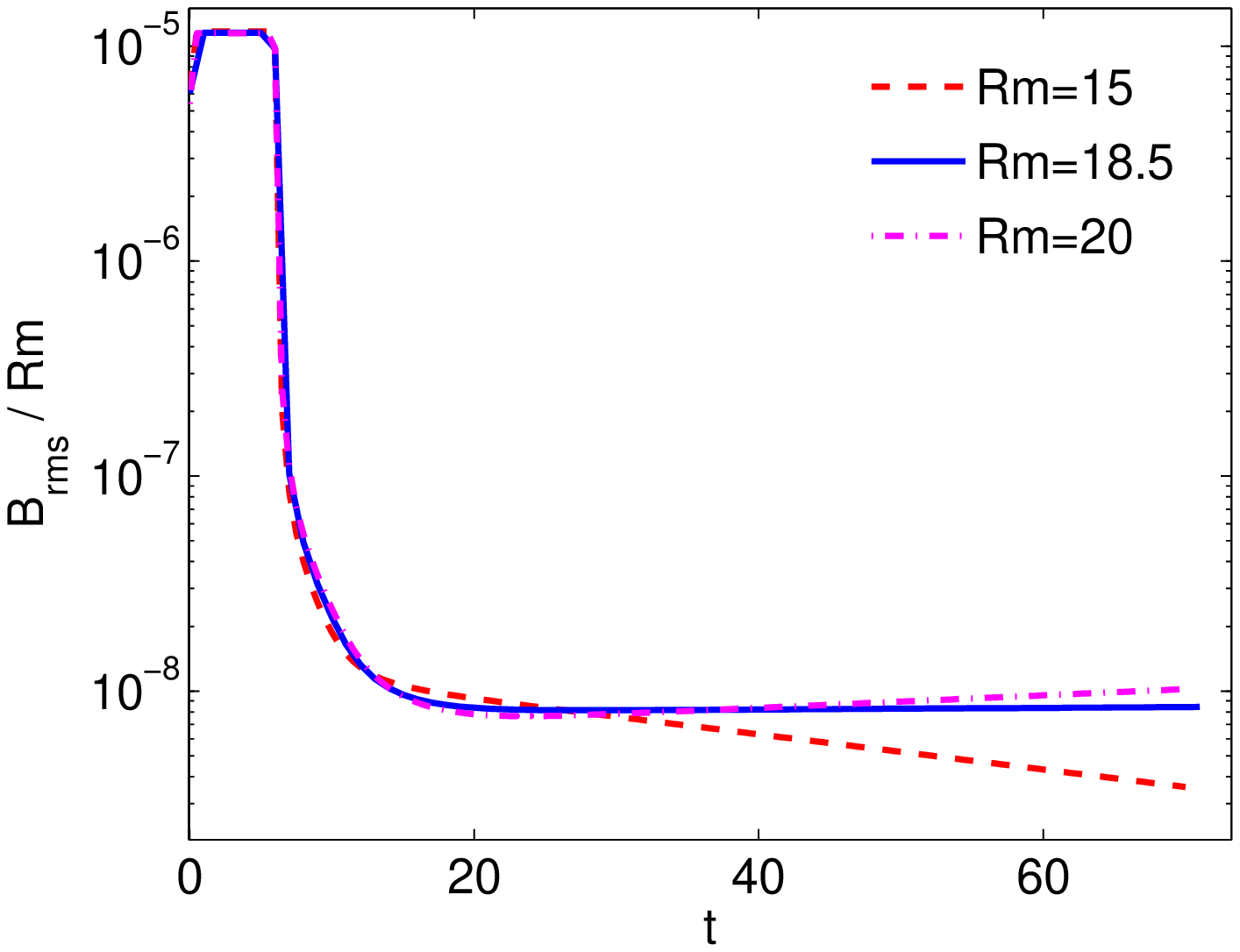}} &
      \setlength{\epsfysize}{6.5cm}
      \subfigure[]{\epsfbox{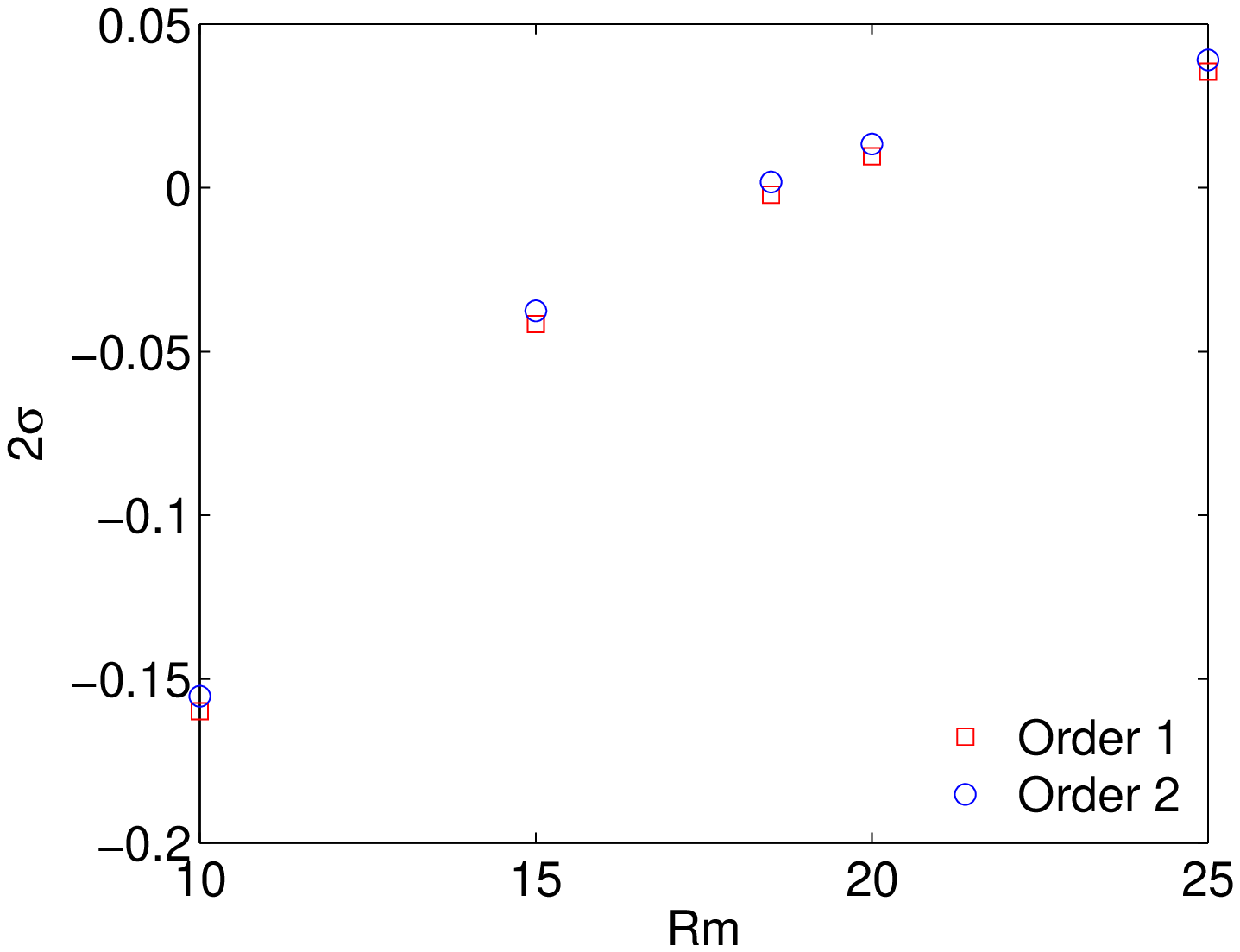}}
    \end{tabular}
    \caption{Numerical results for the Ponomarenko-like kinematic dynamo. (a) Temporal evolution of the quadratic mean magnetic field divided by the magnetic Reynolds number,
    below the dynamo threshold ($Rm=15$), around the threshold ($Rm=18.5$) and above the threshold ($Rm=20$), using quadratic edge elements.
    (b) Evolution of the growth rate of the dynamo with the magnetic Reynolds number using linear and quadratic Nedelec elements.
    The mesh is the same in both cases, with $43\ 162$ tetrahedral elements, but the order of the elements leads to a model with $52239$ DoF using linear edge element,
    and $279560$ DoF using quadratic edge elements. }
    \label{cebronfig3}             
  \end{center}
\end{figure}

A typical result of the magnetic field excited above the dynamo
threshold is shown in figure \ref{cebronfig1}b. This structure is in
perfect agreement with the expected field \cite[see
e.g.][]{Laguerre_2006}. To study more precisely the dynamo
threshold, we consider the temporal evolution of the dimensionless
quadratic mean magnetic field strength
\begin{eqnarray}
B_{rms}=\sqrt{\frac{1}{V_s}\int_{V_s} B^2\ \mathrm{d}V}, \label{eq:Brms}
\end{eqnarray}
where $V_s$ is the volume of the cylinder of radius $1$ non-dimensionalized by
$R^3$. Three examples are shown in figure \ref{cebronfig3}a,
respectively below, above and at the dynamo threshold. The growth/decay rate
$\sigma$ of the magnetic field shown in figure \ref{cebronfig3}b, is
deduced from the growth/decay rate $\sigma$ of $B_{rms}$, determined
from the exponential fit of its temporal evolution. The threshold is
precisely found at $Rm_c \approx 18.3$, in excellent agreement with
the threshold of $Rm_c \approx 18.5$ found by \cite{Laguerre_2006}.
Note that in this case, the order of the edge (N\'ed\'elec) elements
used to compute the magnetic field does not really matter.

\subsection{Von Karman kinematic dynamo: test of the ferromagnetic boundary conditions}\label{valid3}

\begin{figure}                   
  \begin{center}
    \begin{tabular}{ccc}
      \setlength{\epsfysize}{6.5cm}
      \subfigure[]{\epsfbox{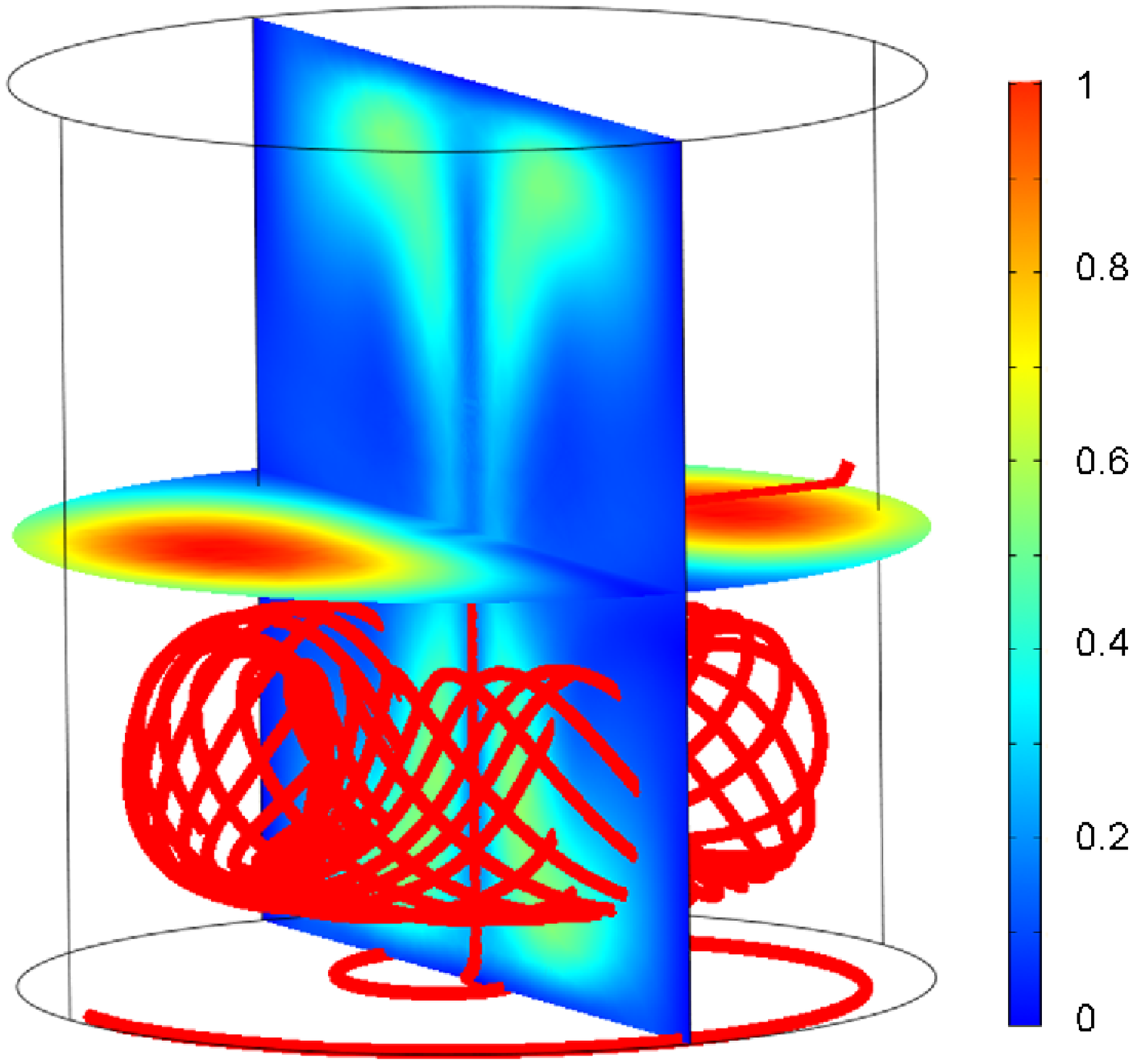}} &
      \setlength{\epsfysize}{6.5cm}
      \subfigure[]{\epsfbox{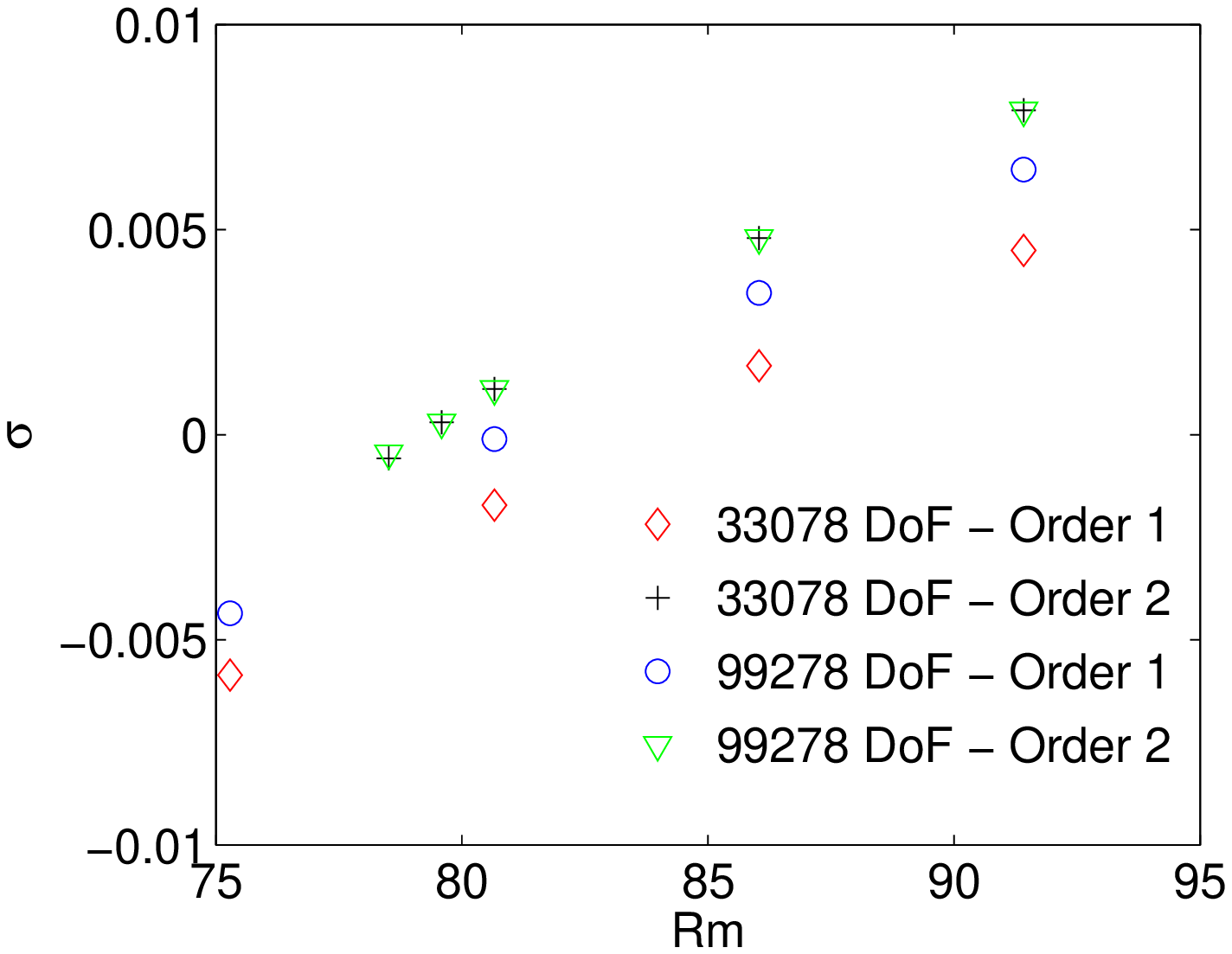}}
    \end{tabular}
    \caption{Numerical results for the Von Karman kinematic dynamo using ferromagnetic boundary conditions. (a) Considering a simulation above the dynamo threshold ($Rm=80.6$, 99278 DoF, using quadratic vector elements), the norm of the magnetic field (normalized by its maximum value) during its exponential growth is represented on slices. Streamlines of the velocity field used, given by (\ref{eq:gissinger}), are also shown (only in the lower half of the cylinder).  (b) The evolution of the growth/decay rate $\sigma$ of the kinematic dynamo is represented as a function of the magnetic Reynolds number for two different meshes, with the linear or the quadratic edge elements. The threshold value $Rm_c=79$ is obtained by interpolation.}
    \label{cebronfig4}             
  \end{center}
\end{figure}

To test the quasi-vacuum boundary condition $\boldsymbol{n} \times
\boldsymbol{B}= \boldsymbol{0}$, we consider a Von Karman kinematic
dynamo in a cylinder of radius $R$ and of aspect ratio $H/R=2$. The
dimensionless base flow is given in \cite{Gissinger_2009}
\begin{eqnarray}
\boldsymbol{U}=
\left [
   \begin{array}{ccc}
      \displaystyle U_r \\[3 mm]
      \displaystyle U_{\theta} \\[3 mm]
      \displaystyle U_z
   \end{array}
   \right ]=
\left [
   \begin{array}{ccc}
      \displaystyle -\frac{\pi}{2}\ r\ (1-r)^2\ (1+2\ r) \cos(\pi z) \\[3 mm]
      \displaystyle \frac{8}{\pi}\ r\ (1-r)\ \arcsin(z) \\[3 mm]
      \displaystyle (1-r)(1+r-5\ r^2)\ \sin (\pi z)\
   \end{array}
   \right ] \label{eq:gissinger}
\end{eqnarray}
Following \cite{Gissinger_2009}, we define here the magnetic
Reynolds number by $Rm=U_{max}\ R/\nu_m$, with $U_{max}$ the peak
velocity of the mean flow. In figure \ref{cebronfig4}a, the magnetic field and the base flow (\ref{eq:gissinger}) are shown. As expected, the magnetic field induced by this flow is an equatorial dipole \cite[e.g.][]{Gissinger_2009}. On the contrary, note that the magnetic field observed in the Von-Karman Sodium experimental dynamo is an axial dipole, which has been attributed to the non-axisymmetric component of the flow  \cite[][]{Gissinger_2009}.
In figure \ref{cebronfig4}b, the
growth/decay rate $\sigma$ for different meshes and orders of
elements are shown. With the second order elements, the results are
converged, and we find a critical magnetic Reynolds number
$Rm_c=79.2$ with $U_{max}=1.0755$, reached in $(r,z)=(0.4842, \pm
1)$, which is in very good agreement with the threshold value $Rm_c=79$
found numerically by C. Nore and A. Giesecke using two other codes
(private communications). On the other hand, this threshold differs
significantly from the value $Rm=60$ given by \cite{Gissinger_2009}.
Figure \ref{cebronfig4}b shows also that the order of elements
has an influence with this kind of boundary conditions even if the
values obtained are close to each others. Actually, a closer look on
the magnetic field at the boundary shows that the boundary
conditions are much more respected with the second order elements
than with the first order ones. Anyway, this validation case shows
that the code is able to reproduce correctly ferromagnetic
conditions.

\subsection{Thermal convection dynamo benchmark}\label{valid4}

As shown in the previous sections, our numerical approach is able to
study kinematic dynamos. The present section extends the validation
to dynamic dynamos considering the usual numerical benchmark of
\cite{Christensen} driven by thermal convection in a rotating
sphere. Up to now, this benchmark, initially defined with spectral methods, has only been considered in two works using local numerical methods: \cite{Matsui_2004,Matsui_2005} reproduced the benchmark on the Earth Simulator with a finite element method based on a potential formulation; \cite{Harder_2005} used a finite volume method and consider a slighty different case, using pseudo-vaccum conditions at the external boundary. In the present work, we consider the case of \cite{Harder_2005}, using our commercial software, based on vector elements. 

We thus solve also the energy equation and add a buoyancy force in
the Navier Stokes equations, as already shown in
\cite{Cebron_2010b}. The Boussinesq approximation is used and
gravity varies linearly with radius. For direct comparison, we use
here the scaling of \cite{Christensen}. The considered geometry
corresponds to a rotating spherical shell of aspect ratio
$\eta=0.35$, with an outer radius $r_0=20/13$ and the inner radius
$r_i=7/13$, where the length scale is the gap $D$. Temperatures are
fixed at $T_o$ and $T_o + \Delta T$ at the outer and inner
boundaries, respectively. The time scale is $D^2/ \nu$, and
dimensionless temperatures are defined as $(T-T_o)/\Delta T$. The
dimensionless temperatures on the outer/inner boundary are thus
equal to $0$ and $1$, respectively. Magnetic induction $B$ is scaled
by $\sqrt{\rho \mu \nu_m \Omega}$, and the pressure by $\rho \nu
\Omega$. Non-dimensional control parameters are the modified
Rayleigh number $Ra=\alpha\ g_0\ \Delta T D/(\nu\ \Omega)=100$,
where $\alpha$ is the thermal expansion coefficient and $g_0$ the
gravity at the outer radius, the Ekman number $E=\nu/(\Omega\
D^2)=10^{-3}$ and the thermal Prandtl number $Pr=1$. Compared to the model introduced in section \ref{sec:num}, the problem is
solved in the frame in rotation with the spherical shell, where the no-slip conditions give a
vanishing velocity on the boundaries. Because of the computational
cost, we use the quasi-vacuum boundary condition $\boldsymbol{n}
\times \boldsymbol{B}= \boldsymbol{0}$ on the outer radius, already
used in this case by \cite{Harder_2005}. This is different from the
benchmark conditions, where a potential magnetic field matching is
used, which means that the obtained solution could be slightly
different. As explained in \cite{Christensen}, because non-magnetic
convection is found stable against small magnetic perturbations at
these parameters and because the dynamo solutions seem to have only
a small basin of attraction, the initial state is of some concern.
We use the same initial conditions as in \cite{Christensen} benchmark. In our case,
the initial condition on the magnetic field has to be written for
the magnetic potential. The calculation of the potential vector $\boldsymbol{A}$ reads:
\begin{eqnarray}
\boldsymbol{A}=\left [
   \begin{array}{ccc}
      \displaystyle A_r \\[3mm]
      \displaystyle A_{\theta}\\[3mm]
      \displaystyle A_{\phi}\
   \end{array}
   \right ]=
\left [
   \begin{array}{ccc}
      \displaystyle \frac{5}{2}\ r\ \sin [\pi(r-r_i)] \cos (2\theta)+f_1(r)+\int(A_{\theta}+r\ \partial_r  A_{\theta}) \mathrm{d} \theta \\[3mm]
      \displaystyle f_2(r,\theta) \\[3mm]
      \displaystyle \frac{5}{8} \left[ 4 r_0 r-3 r^2-\frac{r_i^4}{r^2} \right] \sin \theta+\frac{K}{r \sin \theta}\
   \end{array}
   \right ]
\end{eqnarray}
with the arbitrary functions $f_1(r)$, $f_2(r,\theta)$ and the arbitrary constant $K$. As in \cite{Christensen}, we define the mean magnetic energy density in the shell by
\begin{eqnarray}
E_m=\frac{1}{2\ V_s\ E\ Pm} \int_{V_s} \mathbf{B}^2 \mathrm{d}V
\end{eqnarray}
where $V_s$ refers to the dimensionless volume of the fluid shell.
The mean magnetic field strength $B_{rms}$ used by
\cite{Harder_2005} for this benchmark is an equivalent quantity
defined by $B_{rms}=\sqrt{2\ E_m}$. The dimensionless initial
magnetic energy is then $E_m=868$, i.e. $B_{rms}=41.7$.

\begin{figure}                   
  \begin{center}
    \begin{tabular}{ccc}
      \setlength{\epsfysize}{6.5cm}
      \subfigure[]{\epsfbox{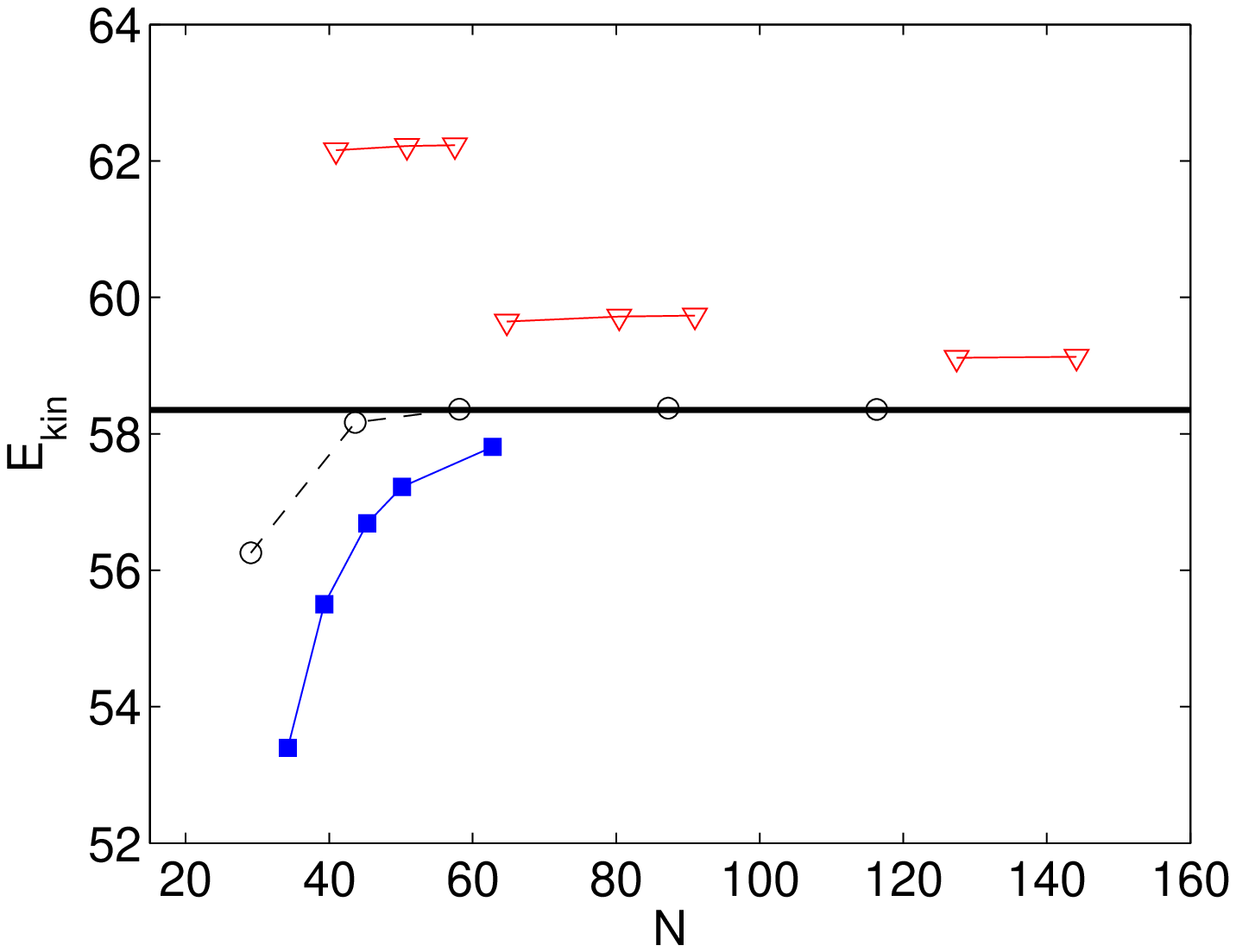}} &
      \setlength{\epsfysize}{6.5cm}
 %     \subfigure[]{\epsfbox{cebronfig77.eps}} &
%      \setlength{\epsfxsize}{3.0cm}
%      \setlength{\epsfysize}{2.5cm}

      \subfigure[]{\epsfbox{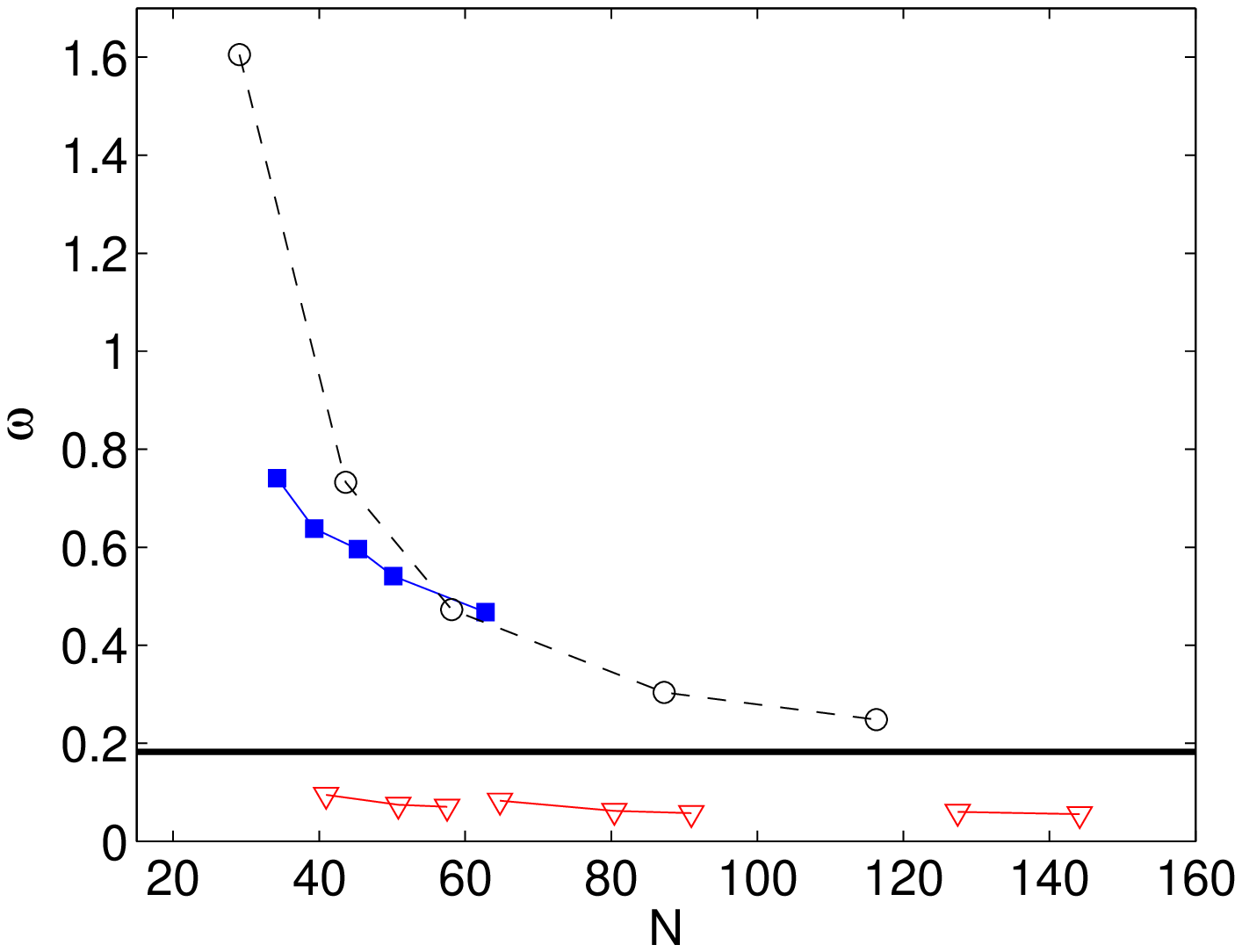}}
    \end{tabular}
    \caption{Evolution of (a) the kinetic energy and (b) the drift angular frequency in our numerical results (squares) for the non-magnetic case of the numerical benchmark of
    \cite{Christensen}, as a function of spatial resolution $N$.
    The results are compared with the suggested solutions (continuous black line) given in \cite{Christensen}, with the results of \cite{Harder_2005} given by the (black) open circles,
    and with the numerical results of \cite{Matsui_2005} given by the (red) triangles. }
    \label{cebronfig5}
  \end{center}
\end{figure}

We first consider the case 0 of the benchmark, which is a
non-magnetic simulation. This allows to check the validity of the
thermally driven flow. The results are shown in figure
\ref{cebronfig5} as a function of the spatial resolution $N$,
defined by the third root of the number of degrees of freedom for
each scalar variable, as in \cite{Matsui_2005}. We focus on the mean
kinetic energy density
\begin{eqnarray}
E_{kin}=\frac{1}{2\ V_s}\ \int_{V_s} u^2\ \mathrm{d} V
\end{eqnarray}
and on the drift frequency $\omega$ of the large-scale convective
columns. As expected and already noticed by \cite{Matsui_2005}, the convergence
is much slower with finite-element methods than with spectral
methods, especially on the drift frequency. However, figure \ref{cebronfig5} shows that our results are
in agreement with those of \cite{Matsui_2005} and \cite{Harder_2005}. Both variables $(E_{kin},\omega)$ converge towards the expected values as $N$ increases.

We now consider the case 1 of the benchmark, solving the full MHD
problem. In this case, the simulations are computationally very
expensive and we are limited to $N \approx 42$ with quadratic N\'ed\'elec elements. Figure
\ref{cebronfig5}a shows that in this case, the mean kinetic energy
density is underestimated of a factor about $5 \%$, which means that the flow is expected to
be less efficient to drive a dynamo. We thus expect a dynamo
threshold a bit larger than in the benchmark. Indeed, we find a dynamo
threshold around $Pm \approx 7$, significantly larger than the
reference value of \cite{Christensen} $Pm=5$. Note nevertheless that
our results are coherent with those of \cite{Harder_2005}, who find
an apparent slowly decaying field for $Pm=5$ (their figure 8) but a
stable dynamo for $Pm=8$. The authors suggest that the slowly decaying field for $Pm=5$ corresponds to the dynamo threshold, and propose to compare the last value reached in their simulation with the benchmark value. They thus compare the mean field strength $B_{rms} \approx
42$ with the mean field strength of $B_{rms}
\approx 35$ given in the benchmark. In our numerical simulation, at
our dynamo threshold around $Pm \approx 7$, we obtain a mean field
strength about $B_{rms} \approx 10$. Once again, this lower value is
expected because our resolution underestimates the kinetic energy of
the flow. With the current numerical power available to us, we are
not yet in a position to go any further. Nevertheless, we are
convinced that the above results are encouraging towards the
validation of our code to solve full dynamo problems.

\section{Application to the MHD elliptical instability} \label{sec:MHD_elliptic}

The next step of our work is in the direct continuity of our
previous numerical studies of the elliptical instability in
non-axisymmetric geometries
\cite[][]{Cebron_2010a,Cebron_2010b,Cebron_2010c}. We consider a
triaxial ellipsoid of axes $(a,b,c)$ with $a>b$, related to the
frame $(Ox,Oy,Oz)$, with an imposed constant tangential velocity
along the deformed boundary in each plane perpendicular to the rotation axis, chosen here as the axis $(Oz)$. Such a configuration is a model for a liquid planetary core with no solid inner core, surrounded by a solid mantle tidally deformed by a companion body. When the differential rotation between the fluid and the deformation is constant, a TDEI can be excited (see section \ref{sec:intro}), as it may be the case for the liquid core of the Early Earth \cite[][]{Cebron_2010a,Cebron_2011}. In this framework, sections \ref{spinover} and \ref{mode13}, which respectively focus on the magnetic induction by a stationary and an unstationary mode of the TDEI, are relevant to the dynamics of the Early Earth liquid core, considering an imposed magnetic field created by an independent convective geodynamo process. When the differential rotation between the fluid and the deformation is oscillatory, a LDEI can be excited as it may be the case for the liquid core of Europa \cite[][]{KerswellMalkus,Cebron_2011}. The case study of the magnetic induction by LDEI presented in section \ref{LDEI} could thus be related to the magnetic induction of a possibly excited LDEI in Europa considering the presence of the jovian magnetic field. Such an internal process could be of fundamental importance to correctly interpret the recorded magnetic data.

In the following, we use the mean
equatorial radius $R=(a+b)/2$ as a length scale and $\Omega^{-1}$ as a time
scale, where $\Omega R$ is the imposed boundary velocity at the
equator. In addition to the already introduced dimensionless numbers
$E$ and $Rm$, two geometrical parameters are necessary to fully
describe the system: the ellipticity $\beta=(a^2-b^2)/(a^2+b^2)$ of
the elliptical deformation and the aspect ratio $c/a$.

\subsection{Spinover induced magnetic field}\label{spinover}

\begin{figure}                   
  \begin{center}
    \begin{tabular}{ccc}
      \setlength{\epsfysize}{6.0cm}
      \subfigure[]{\epsfbox{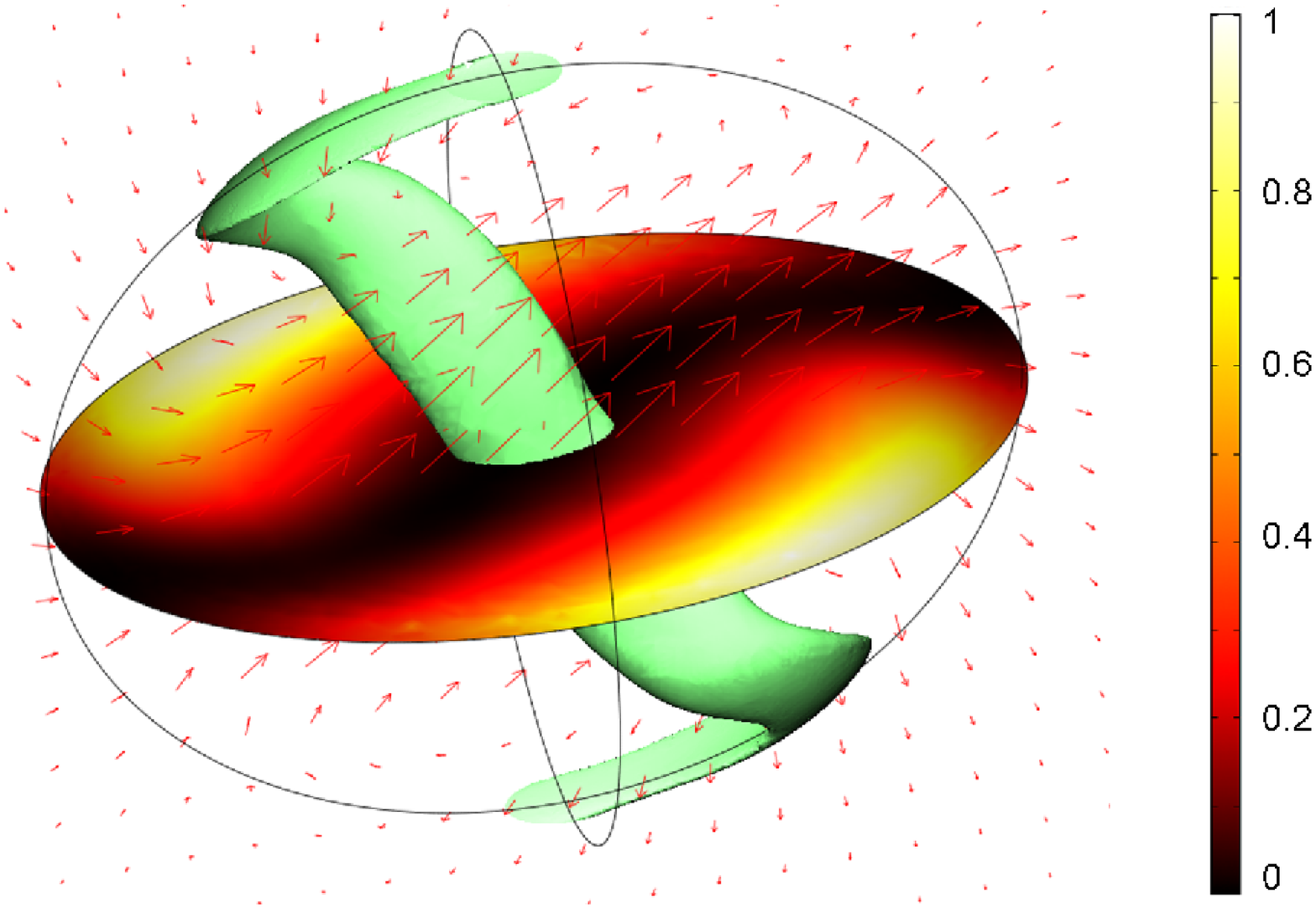}} &
      \setlength{\epsfysize}{6.5cm}
      \subfigure[]{\epsfbox{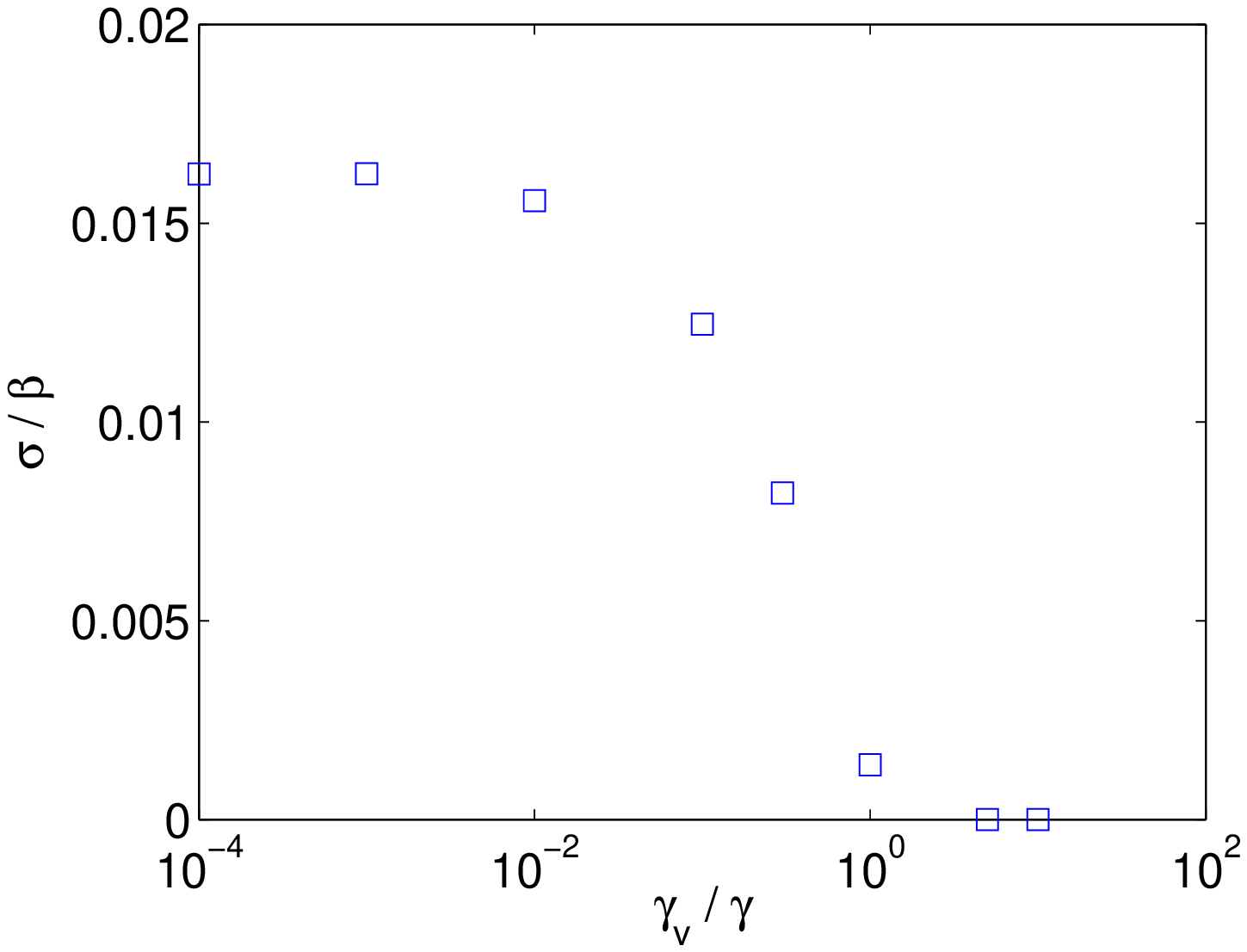}}
    \end{tabular}
    \caption{MHD numerical simulations of the spinover mode in a triaxial ellipsoid, with an uniform magnetic field imposed along the rotation axis.
    The magnetic Prandtl number of the fluid is fixed to $Pm=10^{-4}$, the ellipticity to $\beta=0.317$ and the Elsasser number to $\Lambda=0.02$.
    (a) The spinover mode is shown with an iso-surface of the velocity $||\mathbf{u}||=0.12$, and the induced magnetic fied is represented
    with arrows (the size of the arrows is proportional to the local value of the magnetic field) for $E=1/500$.
    In the equatorial plane, the Joule dissipation is shown, normalized by its maximum value.
    (b) Evolution of the growth rate of the elliptical instability with the ratio between the outer conductivity $\gamma_v$ and the fluid conductivity $\gamma$.}
    \label{cebronfig21}             
  \end{center}
\end{figure}

In this section, the triaxial ellipsoid $(a,b,c)$ is immersed into a
sphere of radius $8\ \sqrt[3]{abc}$ containing a steady material of
electrical conductivity $\gamma_v $. A constant and uniform magnetic
field $\mathbf{B_0}$ is imposed parallel to the rotation axis. We focus first on the
so-called spin-over mode of the elliptical instability (see figure \ref{cebronfig21}a) which is obtained when the length of the polar axis is $c=(a+b)/2$
\cite[see][]{Cebron_2010a}. The magnetic Prandtl number of the fluid
is fixed, $Pm=10^{-4}$. Following \cite{Herreman_2009}, the magnetic
field is non-dimensionalized by the magnetic scale $B_0$, which
simply means that compared to the equations
(\ref{U1})-(\ref{div_free}), the Laplace force in (\ref{U1}) now
writes
\begin{eqnarray}
\frac{\Lambda }{Rm}(\mathbf{\nabla \times B}) \times
\mathbf{B_{tot}}
\end{eqnarray}
with $\mathbf{B_{tot}}=\mathbf{B}+(0,0,1)$ and where the Elsasser
number associated to the imposed magnetic field is defined by
$\Lambda= \gamma\ B_0^2/(\rho \Omega)$. This configuration has
already been studied theoretically and experimentally in
\cite{Lacaze_2006,Thess_2007,Herreman_2009} in the case of an
isolating outer medium. It is explored here for the first time
numerically in an ellipsoidal geometry.

A visual validation is first done on magnetic quantities,
given in figure \ref{cebronfig21}a, in agreement with the
theoretical calculations of \cite{Lacaze_2006}. Figure
\ref{cebronfig21}b shows the influence of the outer conductivity on
the growth rate of the spin-over mode. As expected, for small
conductivity ratios $\gamma_v/ \gamma \lesssim 10^{-3} $, the growth
rate reaches a plateau as $\gamma_v/ \gamma$ decreases: the outer medium behaves as an
insulating medium. Thus, in the following, we use $\gamma_v/\gamma =
10^{-4}$.

Combining the results of \cite{Lacaze_2004} and \cite{Thess_2007},
\cite{Herreman_2009} proposed to model the nonlinear evolution of
the spin-over mode in the laboratory frame of reference by the
nonlinear system
\begin{eqnarray}
\dot{\omega}_x&=&- \alpha_1\ (1+\omega_z)\ \omega_y-(\nu_{so}+\Lambda/4)\ \omega_x, \label{eq:lacaze_eqx} \\
\dot{\omega}_y&=&- \alpha_2\ (1+\omega_z)\ \omega_x-(\nu_{so}+\Lambda/4)\ \omega_y, \label{eq:lacaze_eqy} \\
\dot{\omega}_z&=&\beta\ \omega_x\ \omega_y-\nu_{ec}\ \omega_z+\nu_{nl}\ (\omega_x^2+\omega_y^2) \label{eq:lacaze_eqz}
\end{eqnarray}
where $\boldsymbol{\omega}=(\omega_x(t), \omega_y(t), \omega_z(t))$ is the rotation vector of the spinover
mode, $\alpha_1=\beta/(2-\beta)$ and $\alpha_2=\beta/(2+\beta)$. In
the limit $\beta \ll 1$, the damping terms are known analytically,
as first calculated by \cite{Greenspan}: $\nu_{so}= \alpha \sqrt{E}
= 2.62\ \sqrt{E}$ is the linear viscous damping rate of the spinover
mode, $\nu_{ec}=2.85\ \sqrt{E}$ is the linear viscous damping of
axial rotation and $\nu_{nl}=1.42\ \sqrt{E}$ is the viscous boundary
layer effect on the non-linear interaction of the spinover mode with
itself. The magnetic field only adds a linear term corresponding to
the Joule damping $\Lambda/4$ in the directions perpendicular to the
imposed field. Even if this model does not take into account all the
viscous terms of order $\sqrt{E}$ nor the non-linear corrections
induced by internal shear layers \cite[see][for details]{Lacaze_2004}, it satisfyingly agrees with
experiments, regarding the growth rate as well as the non-linear
saturation of the flow and induced field
\cite[][]{Lacaze_2004,Herreman_2009}.

Linearizing the system around the trivial fixed point
$\boldsymbol{\omega}=\mathbf{0}$, the linear growth rate of the
spin-over mode for $\beta \ll 1$ is given by
\cite[][]{Herreman_2009}
\begin{eqnarray}
\sigma=\frac{\beta}{\sqrt{4-\beta^2}}- \tilde{\nu}_{so}, \label{eq:sig_so}
\end{eqnarray}
where $\tilde{\nu}_{so}=\nu_{so}+\frac{\Lambda}{4}$. Above the
instability threshold given by $\beta/\sqrt{4-\beta^2} \geq \tilde{\nu}_{so}$, a
non-trivial stationary state is reached corresponding to
\begin{eqnarray}
\omega_x&=&\pm\ \sqrt{\frac{\nu_{ec}\ [\sqrt{\alpha_1 \alpha_2}-\tilde{\nu}_{so}] } {\alpha_2 \beta-\nu_{nl}\ [\sqrt{\alpha_1 \alpha_2} +\alpha_2^2 / \sqrt{\alpha_1 \alpha_2}] }}
\approx \pm\ \sqrt{\frac{\nu_{ec}\ [\beta-2\ \tilde{\nu}_{so}] } {\beta^2-2\ \nu_{nl}\ \beta }}, \label{eq:lacaze_omx} \\
\omega_y&=&\mp\ \sqrt{\frac{\nu_{ec}\ [\sqrt{\alpha_1 \alpha_2}-\tilde{\nu}_{so}] } {\alpha_1 \beta-\nu_{nl}\ [\sqrt{\alpha_1 \alpha_2} +\alpha_1^2 / \sqrt{\alpha_1 \alpha_2}] }}
\approx \mp\ \sqrt{\frac{\nu_{ec}\ [\beta-2\ \tilde{\nu}_{so}] } {\beta^2-2\ \nu_{nl}\ \beta }}  \approx \mp\ \omega_x, \label{eq:lacaze_omy} \\
\omega_z&=&\frac{\tilde{\nu}_{so}}{\beta}\ \sqrt{4-\beta^2}-1
\approx \frac{2\ \tilde{\nu}_{so}}{\beta}-1, \label{eq:lacaze_omz}
\end{eqnarray}
where approximations are done assuming $\beta \ll 1$. These expressions allow also to obtain the spin-over mode equatorial amplitude \cite[][]{Herreman_2009}:
\begin{eqnarray}
\Omega_{so}=\sqrt{4\ \frac{\nu_{ec}}{ \beta}\ \frac{\sigma}{\beta-4 \nu_{nl}/\sqrt{4-\beta^2}}}.
\end{eqnarray}

\begin{figure}                   
  \begin{center}
    \begin{tabular}{ccc}
      \setlength{\epsfysize}{6.5cm}
      \subfigure[]{\epsfbox{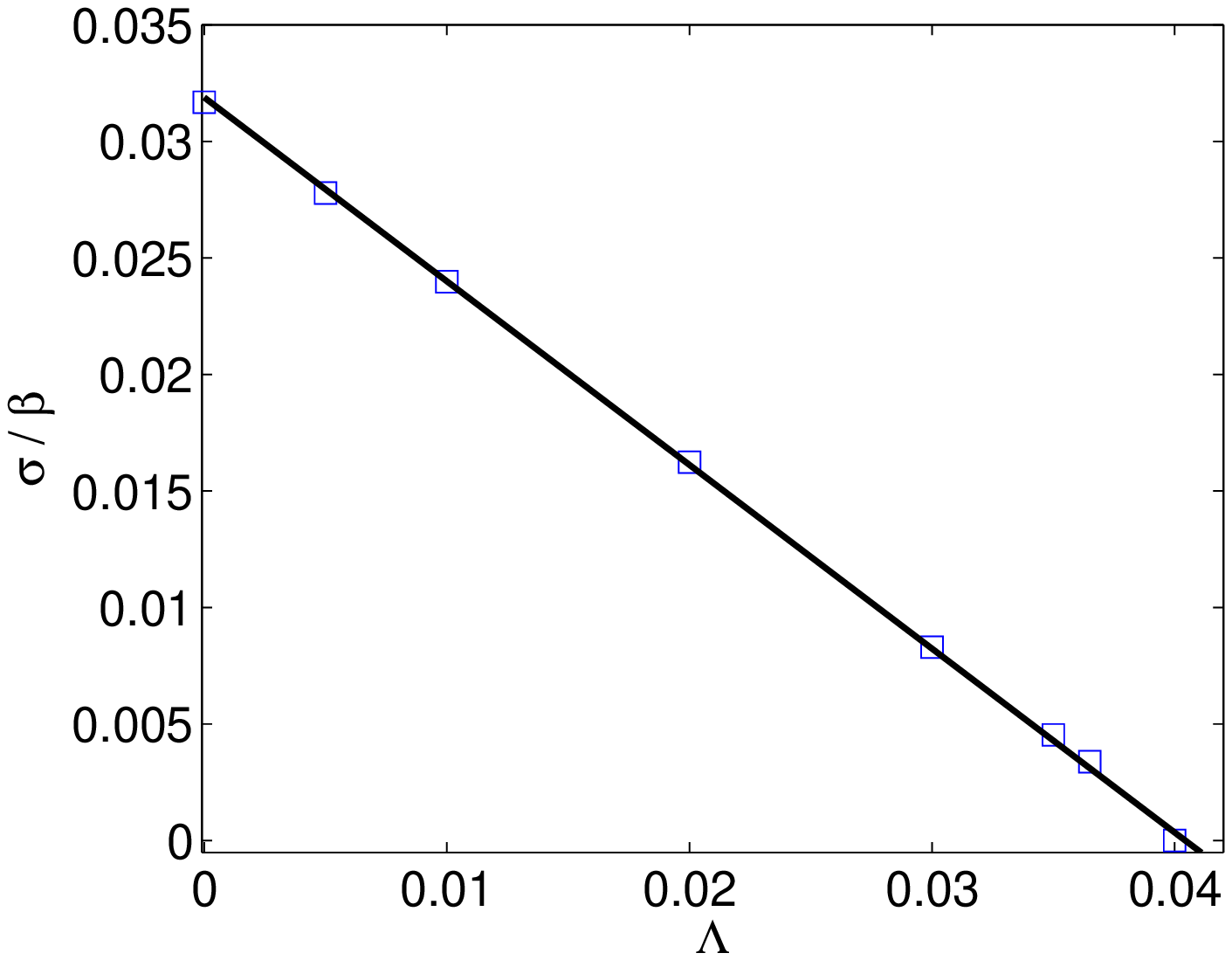}} \\
      \setlength{\epsfysize}{6.5cm}
      \subfigure[]{\epsfbox{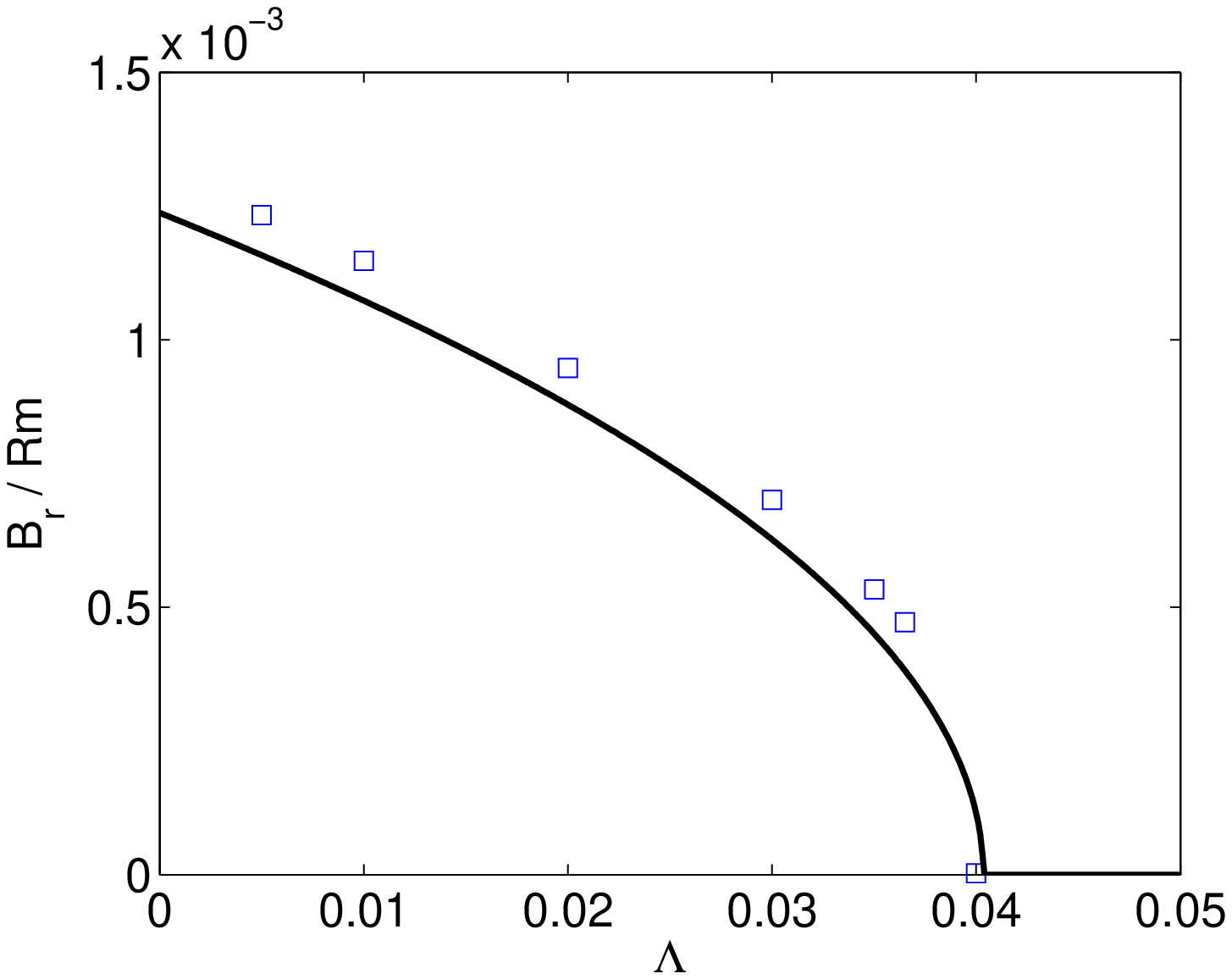}} \\
      \setlength{\epsfysize}{6.5cm}
      \subfigure[]{\epsfbox{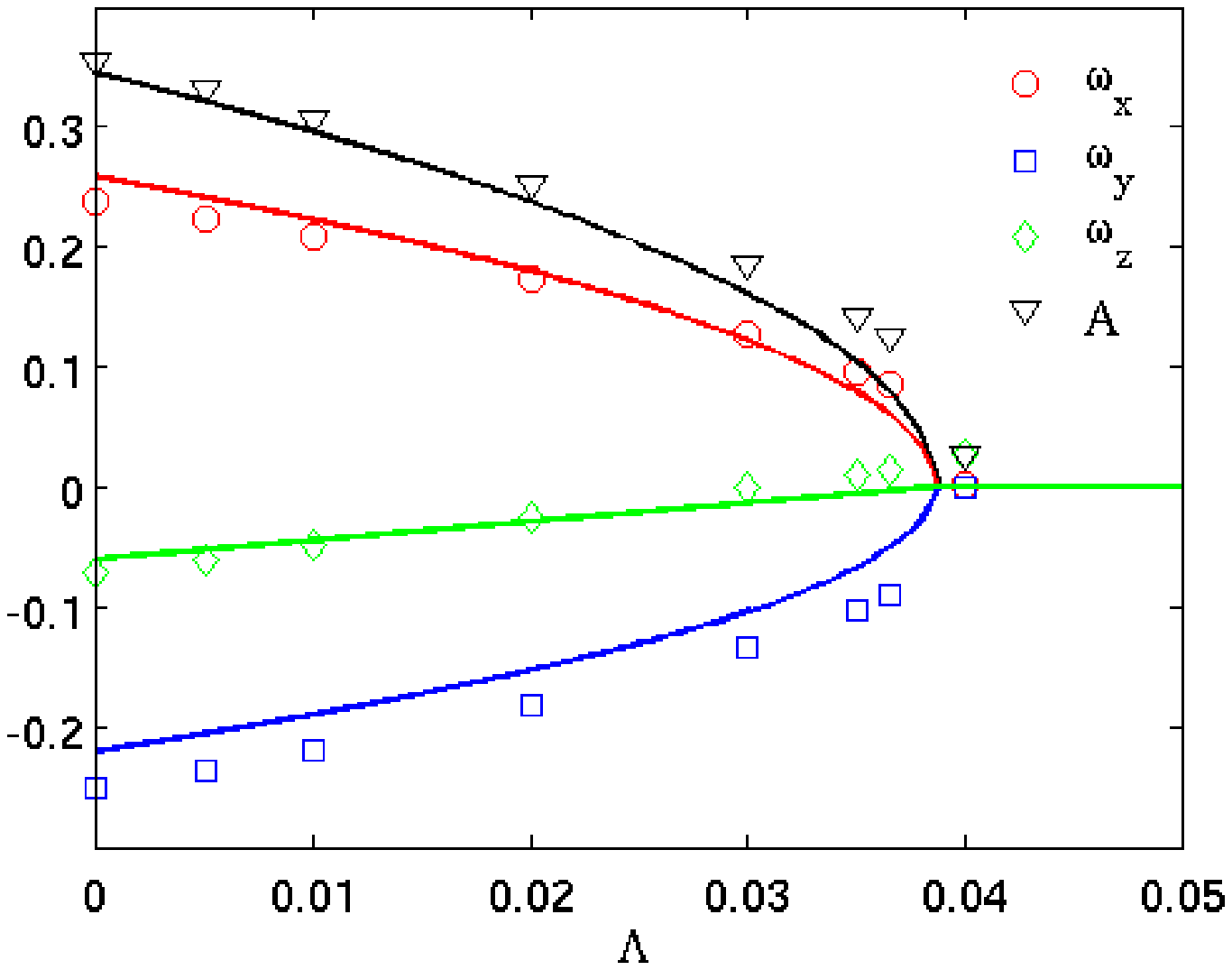}} \\
    \end{tabular}
    \caption{MHD numerical simulations (symbols) and theoretical results (continuous lines) of the spinover mode in a triaxial ellipsoid, with an uniform magnetic field imposed along the rotation axis. Parameters of the simulations are : $E=1/344$, $\beta=0.317$, $c=(a+b)/2$, $Pm=10^{-4}$ for the ellipsoid of fluid, immersed into a sphere of radius $8\ \sqrt[3]{abc}$, with a conductivity $\gamma_v/\gamma = 10^{-4}$. Evolution with the Elsasser number $\Lambda$ of (a) the growth rate of the elliptical instability, (b) the dimensionless amplitude of the induced radial magnetic field at the point of radius $r=2.3$ and longitude $\phi=45^{\circ}$ in the equatorial
    plane ($\theta=\pi/2$) and (c) the components of the spin-over mode solid body rotation. The x-component is represented by red circles, the y-component by blue squares, the z-component by green diamonds and the amplitude $A=||\mathbf{\omega}||$ is given by the black triangles.}
    \label{cebronfig2}
  \end{center}
\end{figure}

According to \cite{Lacaze_2006,Herreman_2009}, the field induced by
the non-viscous spin-over mode at low $Rm$ is a dipole with an axis
transverse to the imposed field, in quadrature with the rotation
axis of the spin-over mode. The axis of length $a$ being the long axis, the polar angle in the $(x,y)$
plane of the saturated spin-over axis can be estimated by
\cite[][]{Herreman_2009}:
\begin{eqnarray}
\phi_{so}= - \left| \textrm{arctan} \left[-\sqrt{\frac{2-\beta}{2+\beta}} \right] \right| \label{eq:phi_so}
\end{eqnarray}
so that the vorticity of the spin-over mode is not exactly aligned
with the direction of the maximum stretching at $- 45^{\circ}$. In
the ellipsoid, the general expression of the induced field is given
in \cite{Lacaze_2006}. In the limit of small magnetic Reynolds
numbers, the cylindrical components $(B_{\rho},B_{\phi},B_z)$ of the
induced magnetic field can be expressed in terms of spherical
variables $(r,\theta,\phi)$ :
\begin{eqnarray}
\mathbf{B_i}(r,\theta,\phi)= Rm\ \Omega_{so} \ \left[
   \begin{array}{ccc}
       \displaystyle  -\frac{1-r^2}{10} \sin(\phi-\phi_{so})+\frac{r^2}{140} (3\ \cos(2 \theta)-1) \sin(\phi-\phi_{so})  \\
       \displaystyle   \frac{1-r^2}{10} \cos(\phi-\phi_{so})+\frac{r^2}{70} \cos(\phi-\phi_{so})   \\
        \displaystyle -\frac{3\ r^2}{140} \sin(2 \theta) \sin(\phi-\phi_{so})  \\
   \end{array}
   \right ],  \label{eq:br_so}
\end{eqnarray}
where $\phi$ is the polar angle in the $(x,y)$ plane,
$r=\sqrt{x^2+y^2+z^2}$ the polar radius. The induced field in the isolating outer domain is then simply given
by $\mathbf{B_e}=r^{-3}\ \mathbf{B_i}(1, \theta, \phi)$.

In figure \ref{cebronfig2}a, the role of the Joule dissipation on
the growth rate of the spin-over mode is studied, comparing the
numerical results with the linear stability solution
$(\ref{eq:sig_so})$. The numerical growth rate is obtained as in
\cite{Cebron_2010a} by a best fit of the initial exponential growth
of the mean amplitude of the vertical velocity $W=1/V_s \cdot
\iiint_{V_s} |w|\ \mathrm{d}V $, with $w$ the dimensionless axial
velocity and $V_s$ the volume of the ellipsoid. The only adjusting
parameter considered here is the viscous damping coefficient
$\alpha$, determined in the absence of magnetic field: we find
$\alpha=2.8$ rather than the theoretical value $\alpha=2.62$,
probably because of the finite value of the ellipticity $\beta$ in
our simulations. No adjustable coefficient is introduced for the
magnetic dependence of the growth rate. The accuracy of the
numerical solution is further tested on figure \ref{cebronfig2}b,
where the evolution of the amplitude of the radial field
$\mathbf{Br}$ with the Elsasser number is compared with theoretical equation
(\ref{eq:br_so}). We have checked that the magnetic field is radial, as expected. Finally, a last quantity of interest is the
amplitude of the flow driven by the instability at saturation. Note
that this quantity is not easily accessible in experiments, and was
not tested in previous studies \cite[][]{Lacaze_2006,Herreman_2009}.
In the numerical simulations, the amplitude $A$ of the flow driven
by the instability at saturation is obtained by determining the mean
additional vorticity of the flow in the bulk of the fluid (i.e.
outside the viscous Ekman layers), in comparison with the imposed
vorticity $2\boldsymbol{e}_z$ due to the imposed rotation. This
value is then compared with the spin-over rotation
vector given by (\ref{eq:lacaze_omx})-(\ref{eq:lacaze_omz}). Results
are shown in figure \ref{cebronfig2}c for the three components of the spin-over mode and the amplitude $A=||\mathbf{\omega}||$. The excellent agreement
exhibited in the three tests presented in figure \ref{cebronfig2}
demonstrates that our numerical model correctly simulates both the
induced field and its retroaction on the flow. We are now in a
position to go further in studying induction by more complex
elliptically driven flows, as relevant for planetary applications
\cite[][]{Cebron_2011}. Note that these complex flows are not easily
accessible to MHD theoretical or experimental approaches.

\subsection{Induced magnetic field by the mode (1,3) of the elliptical instability}\label{mode13}

\begin{figure}                   
  \begin{center}
    \begin{tabular}{ccc}
      \setlength{\epsfysize}{6.3cm}
      \subfigure[]{\epsfbox{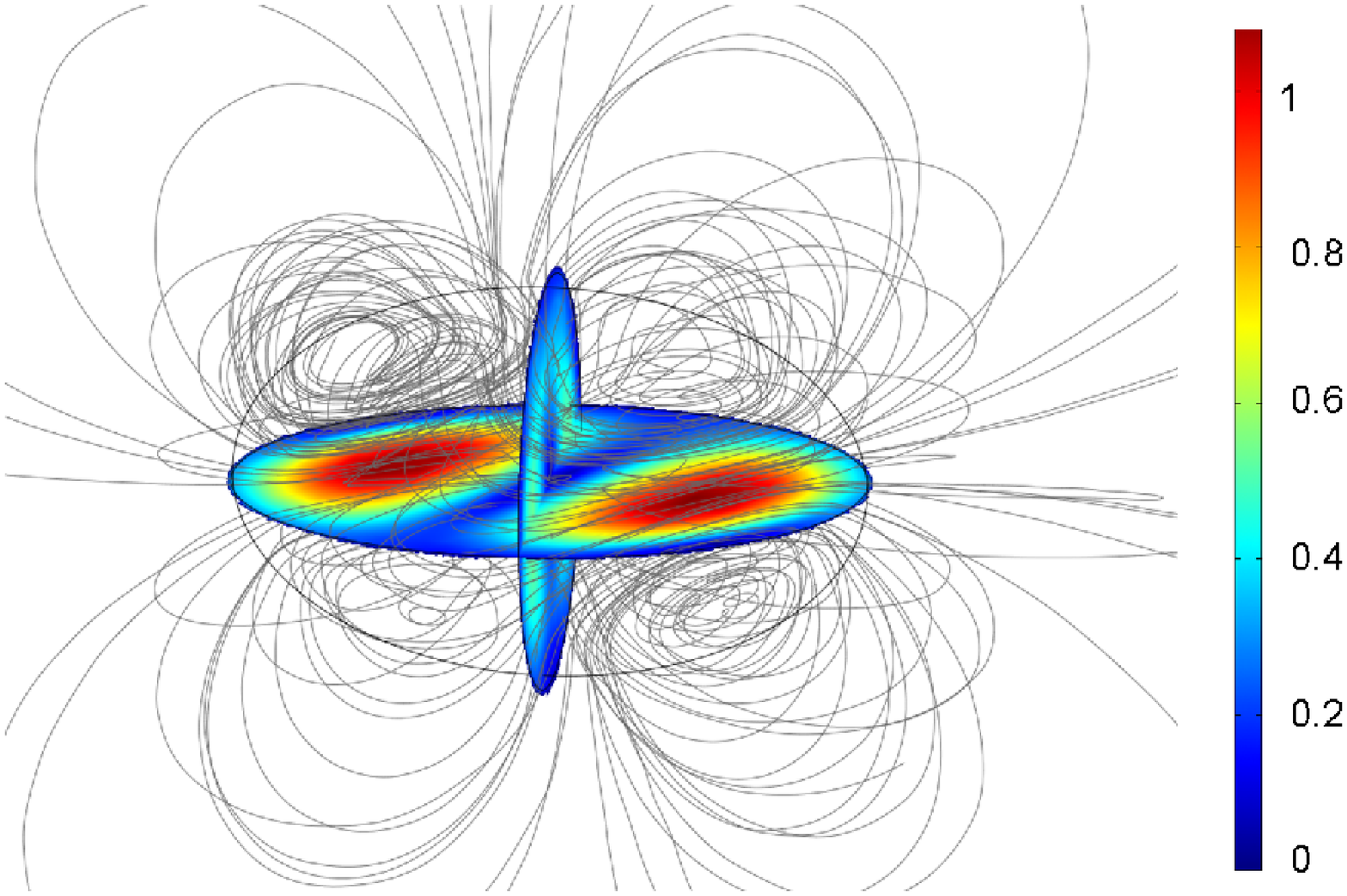}} &
      \setlength{\epsfysize}{6.0cm}
      \subfigure[]{\epsfbox{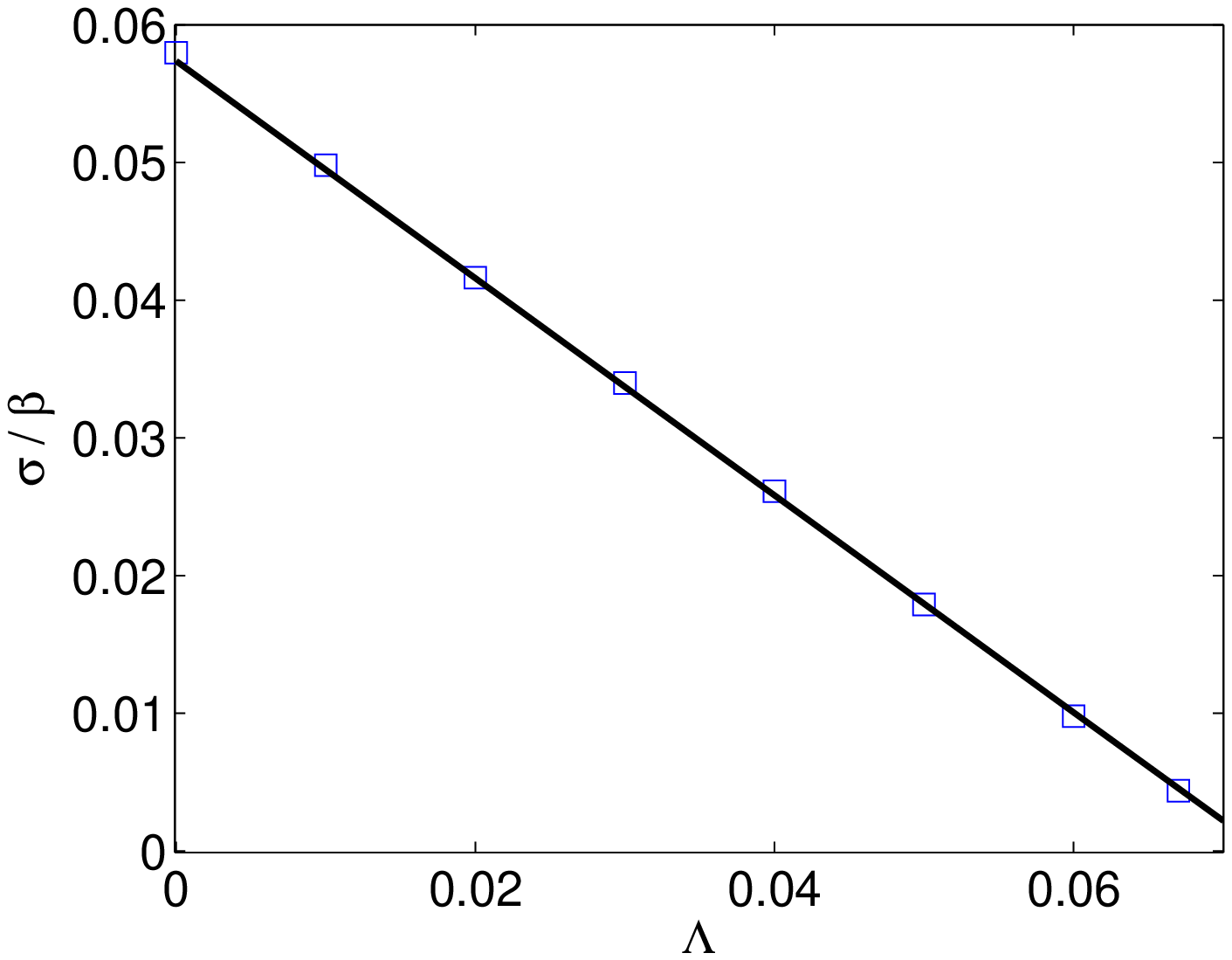}} 
    \end{tabular}
    \caption{MHD numerical simulations of the mode (1,3) of the elliptical instability in a triaxial ellipsoid, with an uniform magnetic field imposed along the rotation axis.
(a) The norm of the induced magnetic field, normalized by its maximum value $B=6.9 \cdot 10^{-4}$, is represented on slices at for the mode (1,3). Magnetic field lines are also shown. Parameters : $\beta=0.317$, $E=1/700$, $c/a=0.65$ and $\Lambda=0.02$.  (b) Evolution of the growth rate of the elliptical instability with the Elsasser number $\Lambda$. The numerical simulations (blue squares) are in
    perfect agreement with the linear stability analysis (continuous black line) given by (\ref{eq:sig_13}), using $\alpha=4.24$, determined in the absence of magnetic field.}
    \label{cebronfig71}
  \end{center}
\end{figure}

Apart from the spin-over mode, no theoretical global approach has
yet been developped for other modes of the elliptical instability.
Our only theoretical tool is then a WKB analysis, where
perturbations of the base field are searched in the form of plane wave
solutions in the limit of large wavenumbers $k \gg 1$ and for $\beta
\ll 1$. Assuming a Laplace force of order
$\beta$, the WKB analysis \cite[][]{Herreman_2009,Cebron_2011}
indicates, in the case of a stationary elliptical distortion, a growth rate
\begin{eqnarray}
\sigma=\frac{9}{16}\ \beta- \alpha \sqrt{E}-\frac{\Lambda}{4}, \label{eq:sig_13}
\end{eqnarray}
where $\alpha$ is again a viscous damping coefficient of order $1$,
and the induced magnetic field $\mathbf{B}$ is linked with the typical velocity of the excited mode $\mathbf{u_0}$ by:
\begin{equation}
 \mathbf{B}= \textrm{i}\ \frac{Rm}{2\ k}\ \mathbf{u_0}, \label{MHD_eq}
\end{equation}
where $k$ is the norm of the wave vector of the excited mode \cite[see][for details]{Cebron_2011}. This
generic expression shows that the induced magnetic field due to the
elliptical instability is systematically proportional to and in
quadrature with the velocity field due to the instability. Note that
both solutions (\ref{eq:sig_13}) and (\ref{MHD_eq}) are in agreement
with the analytical results already obtained by a global method for
the particular case of the spin-over mode (see section
\ref{spinover}).

\begin{figure}                   
  \begin{center}
    \begin{tabular}{ccc}
      \setlength{\epsfysize}{6.48cm}
      \subfigure[]{\epsfbox{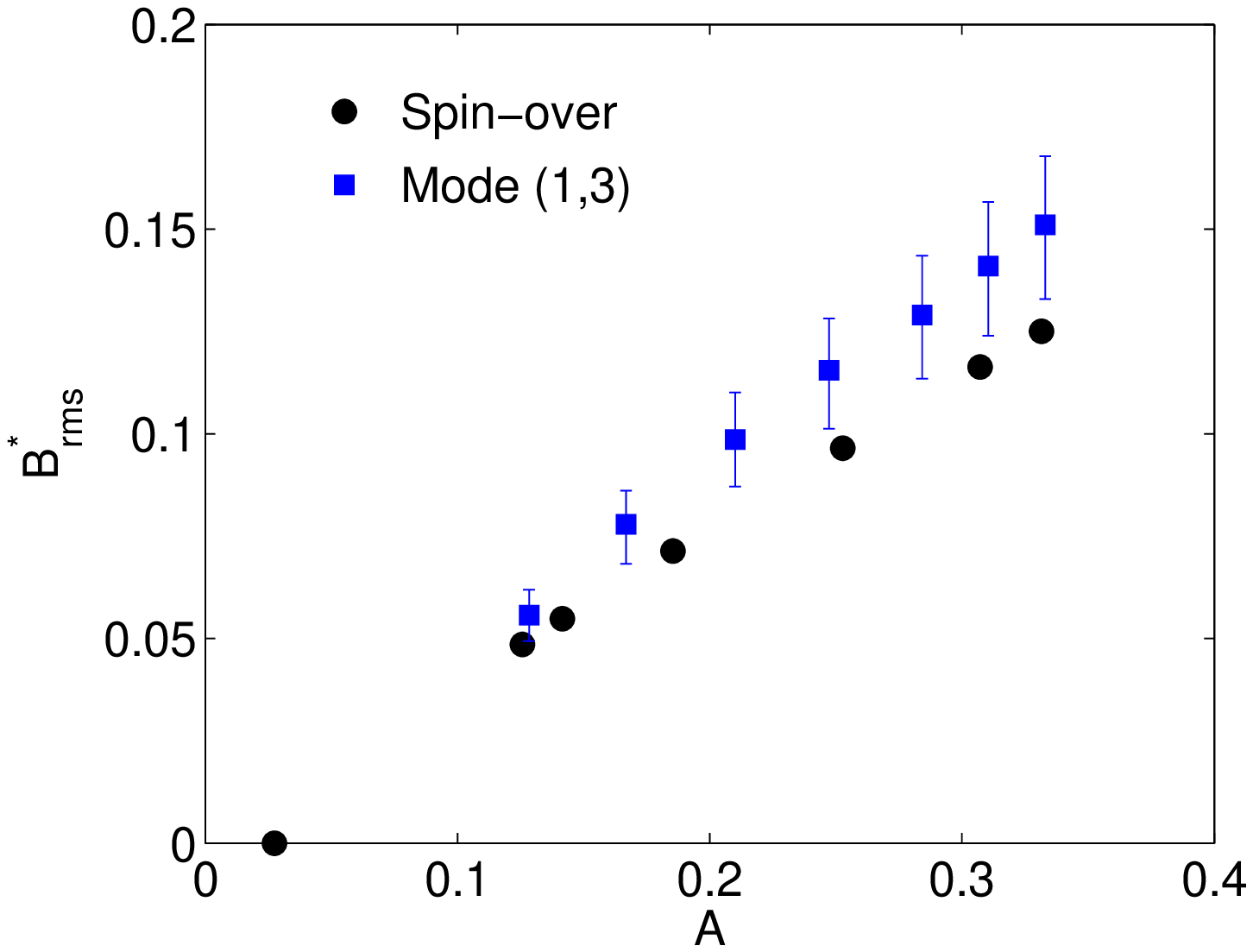}} &
      \setlength{\epsfysize}{6.48cm}
      \subfigure[]{\epsfbox{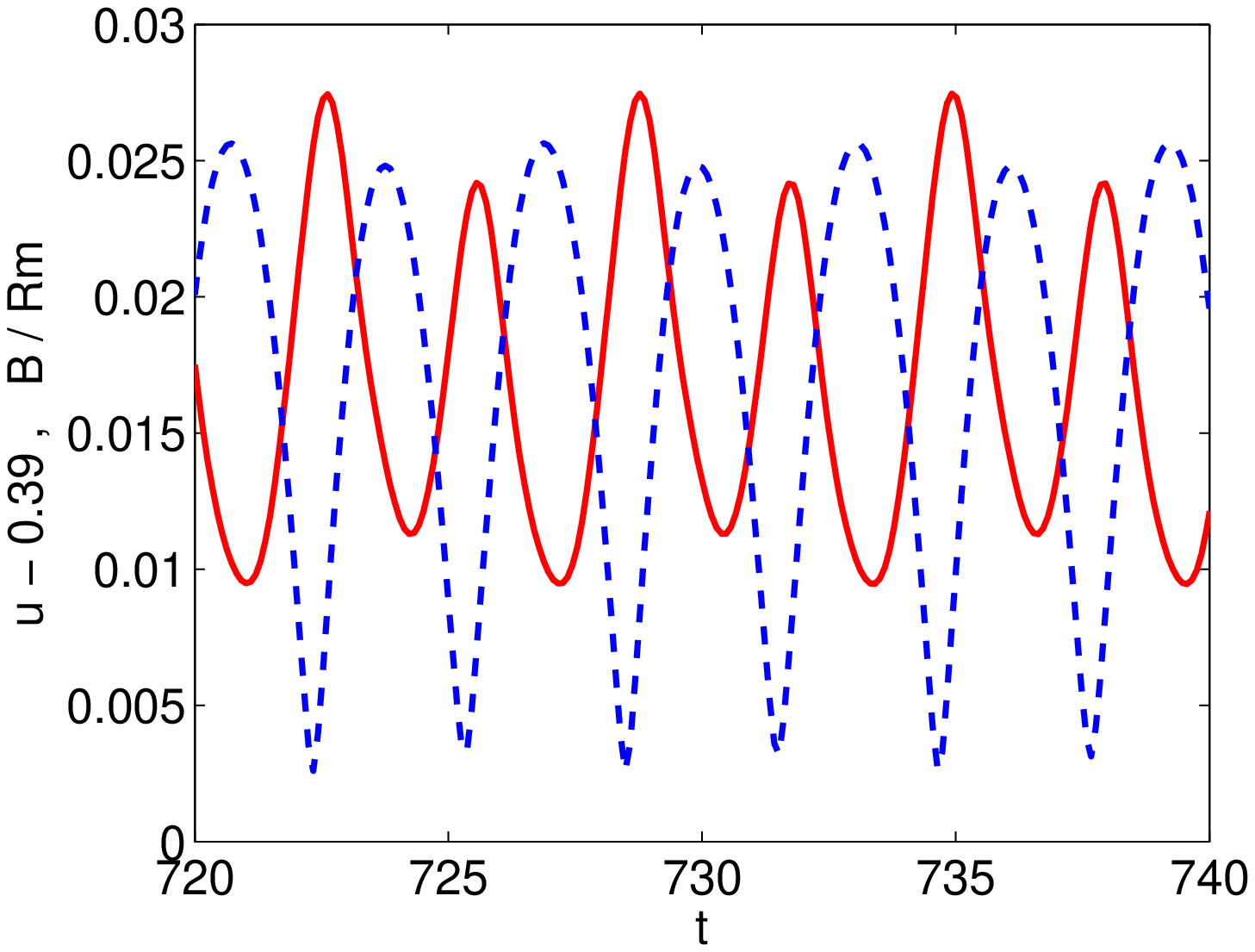}}
    \end{tabular}
    \caption{(a) Evolution of $B_{rms}^*$ with the mode amplitude $A$. For the mode (1,3), the errorbars indicate the extrema values reached by $B_{rms}^*$, showing that the amplitude of the oscillations of the induced magnetic field increases with the distance to the threshold. (b) Time evolution of the velocity quantity $||\mathbf{u}||-0.39$ (shifted for the sake of comparison) and the magnetic field $||\mathbf{B}|| /Rm$ at the point at half the long axis in the equatorial plane. The phase angle shift obtained is around $1.13$, close to the expected value $\pi/2$.}
    \label{cebronfig72}
  \end{center}
\end{figure}

We validate here these results by considering the so-called $(1,3)$
mode of the elliptical instability, which is oscillating $2$ times
faster than the rotation rate of the flow. To do so, the length $c$
of the ellipsoid is fixed to $c/a=0.65$ \cite[see][for
details]{Cebron_2010a}. The typical magnetic field induced by the mode (1,3) is represented in figure \ref{cebronfig71}a. In figure \ref{cebronfig71}b, the excellent
agreement between the theoretical and numerical growth rates of the
instability confirms the general validity of the Joule damping
$-\Lambda/4$. The viscous damping coefficient is found to be equal
to $\alpha=4.24$ in our simulations, which is in the expected range. The expression (\ref{MHD_eq}) is compared with the numerical data in
figure \ref{cebronfig72}a on both the spin-over mode and the mode
(1,3). We defined $B_{rms}^*=2\ k\ B_{rms}/Rm$, where $B_{rms}$ is the quadratic mean value of the magnetic field defined by (\ref{eq:Brms}), $k=2
\pi/\lambda$ and where the wavelength $\lambda$ is equal to
$\lambda=2$ for the spin-over mode and $\lambda=1$ for the mode
(1,3). The typical velocity $u_0$ corresponds to the amplitude $A$ of the $(1,3)$ mode, which is determined by $A=\displaystyle < \max_{V} ||\mathbf{u} - \mathbf{u_b} || >$, where the brackets indicate an average on time and $\mathbf{u_b}$ is the base flow before the destabilization of the elliptical instability. This method is less accurate than the method used in the section \ref{spinover} for the spin-over but is more generic because it can be used for any excited mode. The collapse of the numerical induced
magnetic fields $B_{rms}^*$ for the spin-over mode and the mode
(1,3) at a same distance from the threshold, and the very close
values of the amplitude of the flow and the magnetic field in both
cases, confirm the validity of (\ref{MHD_eq}). Finally in figure
\ref{cebronfig72}b, the phase angle shift suggested by (\ref{MHD_eq}) between
the velocity and the induced magnetic field is validated.

\subsection{Induced magnetic field by the libration driven elliptical instability}\label{LDEI}

\begin{figure}                   
  \begin{center}
      \setlength{\epsfxsize}{7.0cm}
      \epsfbox{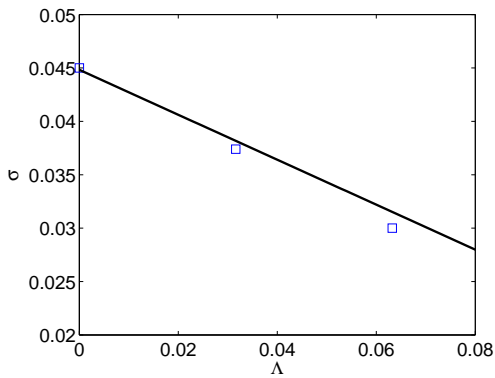}
    \caption{Evolution of the growth rate of the LDEI with the Elsasser number $\Lambda$ associated to the uniform magnetic field imposed along the rotation axis for $\varepsilon=1$, $\omega=1.835$, $\beta=0.44$, $c=1$ and $E=5 \cdot 10^{-4}$. The numerical simulations (blue squares) are in agreement with the linear stability analysis (continuous black line) given by (\ref{eq:sigLDEI}), using $\alpha=3.95$, determined in the absence of magnetic field. }
    \label{cebronfig8}             
  \end{center}
\end{figure}

A recent paper by \cite{Noir_2011} shows the apparition of the
elliptical instability in a librating rigid triaxial ellipsoid, i.e.
in the case where the whole ellipsoid is rotating at a modulated
angular rate $\Omega(t)=\Omega + \Delta\phi f \sin (f t)$. Here,
$\Delta\phi $ is the angular amplitude of libration in radians and
$f$ is the angular frequency of libration. Using the mean equatorial
radius $R$ as the length scale and $\Omega^{-1}$ as the time scale,
the dimensionless angular rate reads $1+\varepsilon \sin(\omega t)$,
with the dimensionless libration frequency $\omega=f/\Omega$ and the
forcing parameter $\varepsilon=\Delta\phi\ \omega$. Note that this
situation is reminiscent of the flow dynamics in the core of
synchronized bodies on average, such as for instance the galilean moons Europa
and Io. As demonstrated theoretically in \cite{Cebron_2011}, a
libration driven elliptical instability (LDEI) grows in certain
ranges of libration frequencies. In presence of an imposed magnetic
field $\mathbf{B_0}$ parallel to the rotation axis, this analysis
gives the theoretical growth rate of the LDEI in the limit of large
wavenumbers $k \gg 1$ and for $\beta , \varepsilon \ll 1$
\begin{eqnarray}
\sigma &=& \frac{16+\omega^2}{64}\ \varepsilon \beta - \alpha \sqrt{E}-\frac{\omega^2}{16}\ \Lambda, \label{eq:sigLDEI}
\end{eqnarray}
with the Elsasser number $\Lambda= \gamma\ B_0^2/(\rho \Omega)$ and
a viscous damping coefficient $\alpha \in [1; 10]$. In an astrophysical context, the Joule damping term can
significantly modify the stability property of the flow, and the
induced field can participate in the magnetic fluctuations measured
for instance in the vicinity of Europa, which is probably the most unstable of the jovian moons  \cite[][]{Cebron_2011}

Our purpose here is to validate these theoretical
predictions. To do so, we use the hydrodynamic numerical model first
presented in \cite{Noir_2011}. Note in particular that we work in
the frame rotating with the ellipsoid, which leads to add the so-called Poincar\'e force to the Navier-Stokes equation (\ref{U1}). In addition to this previous
study, a uniform magnetic field is imposed along the rotation axis,
whereas the ellipsoid is immersed into a sphere of radius $6\
\sqrt[3]{abc}$ where the motionless medium is $10^{-4}$ times less
electrically conductive than the fluid. As in section
\ref{spinover}, we use a magnetic Prandtl number $Pm=10^{-4}$. In
order to favorize the apparition of the LDEI, we choose
$\varepsilon=1$, $\omega=1.835$, $\beta=0.44$, $c=1$, and the Ekman
number $E=\nu/(\Omega_0 R^2)$ is fixed at $E=5 \cdot 10^{-4}$
\cite[see][]{Noir_2011}. The expression of the growth rate
(\ref{eq:sigLDEI}) shows that a small Ekman number is needed to
reach the threshold, because the destabilizing term $\varepsilon
\beta$ is of order two. The increased computational cost imposed by
the value of the Ekman number allows us to perform only three
simulations, for $\Lambda =0;\ 0.032;\ 0.063$. The excited mode of the LDEI in these simulations appears to be a spin-over mode with an oscillating direction. The obtained numerical growth rates are shown in figure \ref{cebronfig8} and confirm the validity of the Joule damping term. Concerning the magnetic field strength, at the point of radius $r=2$ and longitude $\phi=45^{\circ}$ in the equatorial plane ($\theta=\pi/2$), already considered in figure \ref{cebronfig2}b, the magnetic field is radial and equal to $B_r=0.0013$ for $\Lambda =0.032$, and  $B_r=0.001$ for $\Lambda =0.063$. Considering that the theory presented in section \ref{spinover} is still valid for this slightly oscillating spin-over, the theoretical values obtained for the same parameters $\beta$ and E are respectively $B_r=0.0009$ and $B_r=0.0008$, which are close to the numerical values.

\subsection{Dynamo problem}\label{dynamo}

\begin{figure}                   
  \begin{center}
    \begin{tabular}{ccc}
      \setlength{\epsfysize}{6.2cm}
      \subfigure[]{\epsfbox{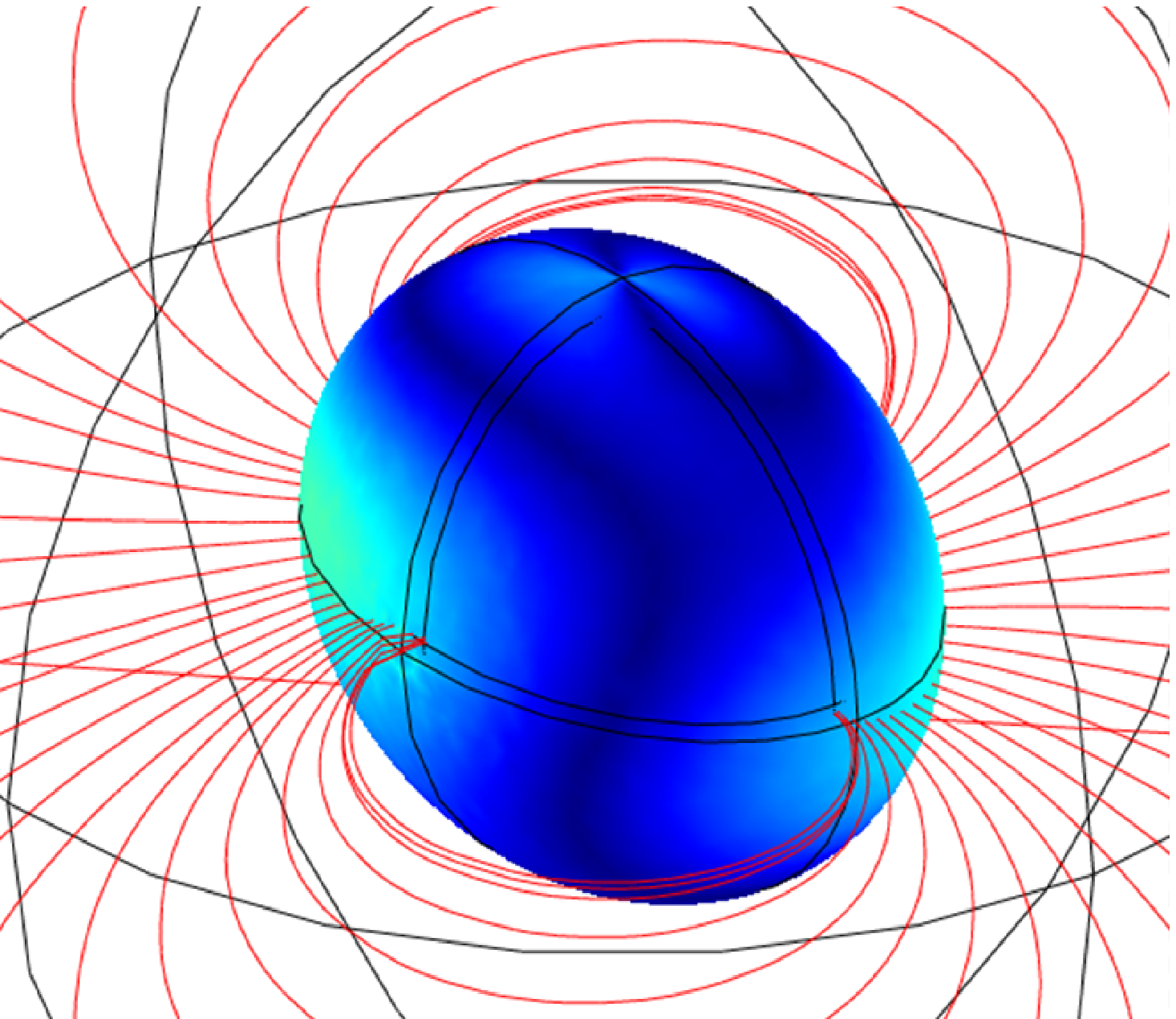}}\\
      \setlength{\epsfysize}{6.8cm}
      \subfigure[]{\epsfbox{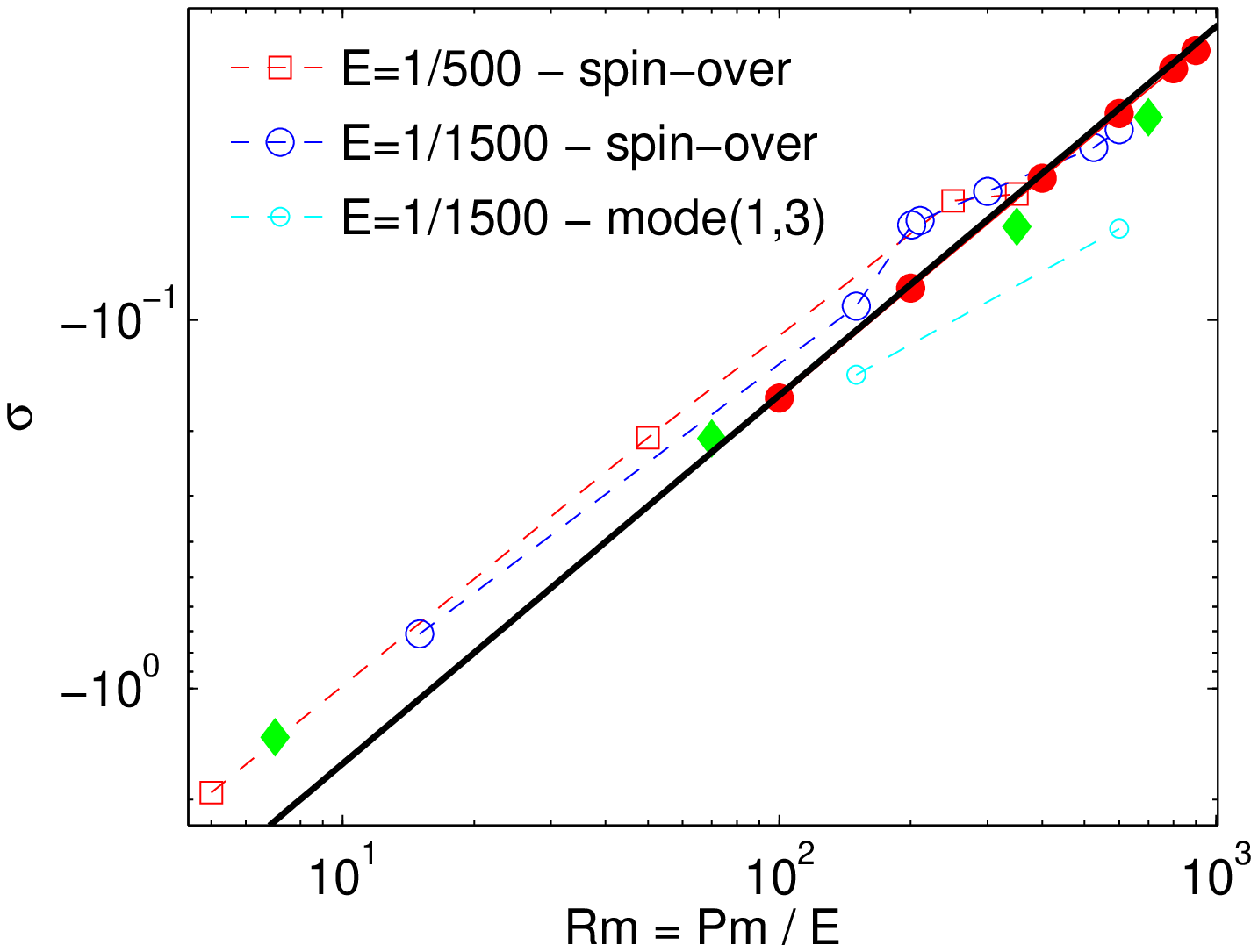}} \\
      \setlength{\epsfysize}{6.8cm}
      \subfigure[]{\epsfbox{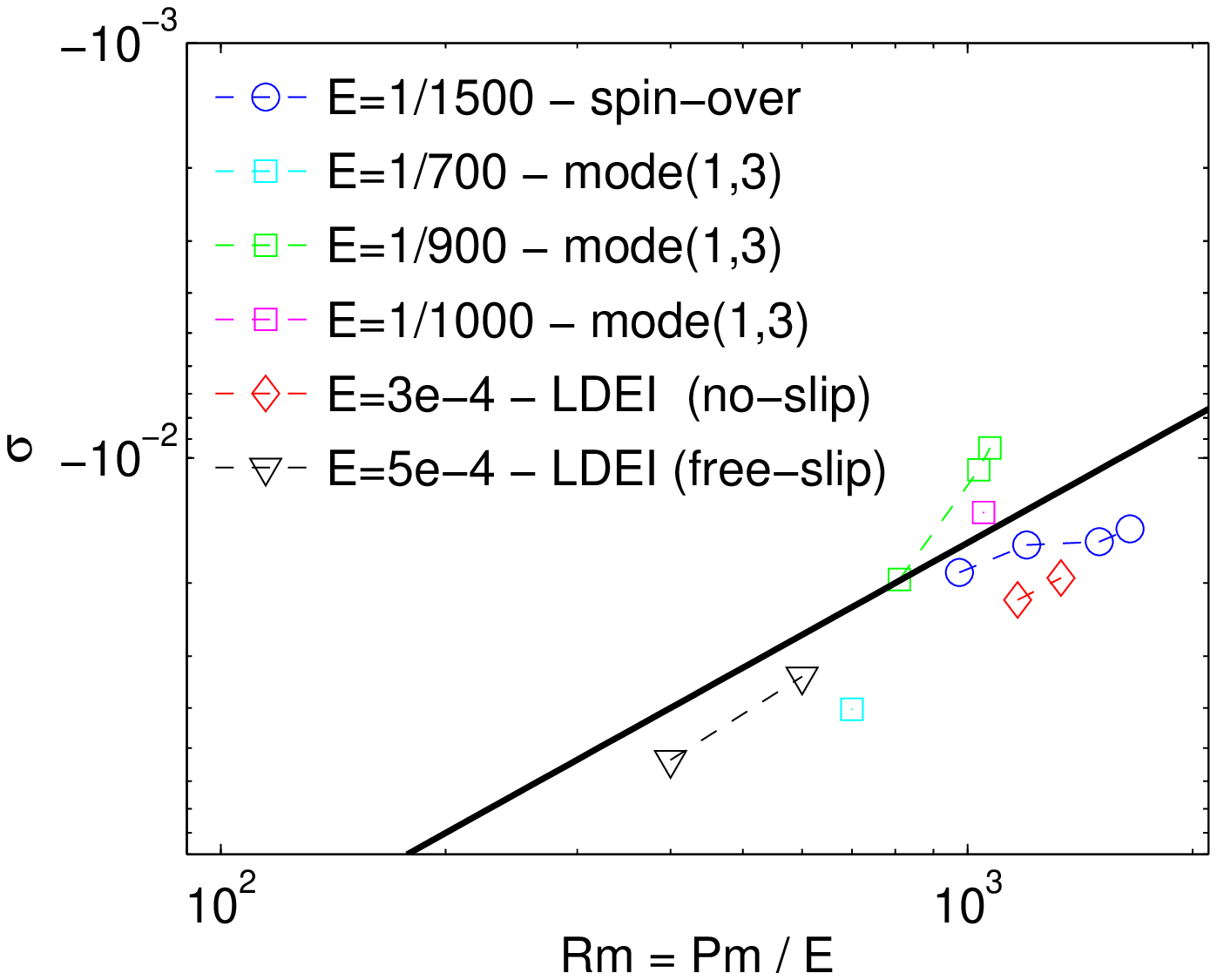}} 
    \end{tabular}
    \caption{Decaying magnetic field once the external imposed magnetic field is shut down.  (a) The norm of the cylindrical outward radial component of the magnetic field (normalized by its maximum value) is represented at the outer boundary of the ellipsoidal fluid domain. This decaying field is obtained for the spin-over mode at the parameters $\beta=0.317$, $E=1/500$, $c=(a+b)/2$ and $Rm=500$. (b) The fluid ellipsoid is immersed into a sphere of radius $10\ \sqrt[3]{abc}$ with an electrical conductivity $10^{-4}$ smaller than the conductivity of the fluid. The figure shows the result for the elliptical instability (open symbols) and for the precession case (solid symbols). The black continuous line stands for the scaling law $-16/Rm$, i.e. a purely diffusive behaviour. The results of \cite{Tilgner_1998} are represented by red circles whereas the green diamonds are the numerical results of our full MHD simulations of a sphere precessing at an angle of $60^{\circ}$, a precession rate $\Omega_p=0.3$ and $E=1/700$. (c) Same as figure (b) but with pseudo-vacuum conditions at the outer boundary. For comparison, some simulations with the LDEI flow are also reported in the case of no-slip boundaries, for $\varepsilon=0.92$, $\omega=1.76$, $\beta=0.44$, $c=1$ and $E=5 \cdot 10^{-4}$; and in the case of free-slip boundaries for $\varepsilon=1$, $\omega=1.8$, $\beta=0.44$, $c=1$ and $E=3 \cdot 10^{-4}$.}
    \label{cebronfig9}
  \end{center}
\end{figure}

Following our induction studies, the next step of our numerical study is to
determine whether or not the elliptical instability is dynamo
capable. To answer this question, and starting from an induction
configuration, we can suddenly shut down the externally imposed
magnetic field and report the decay/growth rate of the induced
magnetic field as a function of the magnetic Reynolds number. Figure \ref{cebronfig9} shows our first numerical results. We are restricted to magnetic Reynolds
numbers $Rm=Pm/E$ lower than $1000$. Indeed, the mesh needed to solve higher magnetic Reynolds numbers leads to a large computational cost, up to now inaccessible. Figure \ref{cebronfig9}a shows the typical dipolar decaying magnetic field for the spin-over mode. Figure \ref{cebronfig9}b shows the systematic report of the decay rate for the spin-over mode and the mode (1,3), at various Ekman numbers, using an insulating outer medium (ratio of conductivities $\gamma_v / \gamma \leq 10^{-6}$). The collapse of the points along the scaling law $Rm^{-1}$ shows that the relevant control parameter is as expected the magnetic Reynolds number. This indicates also that a purely diffusive behaviour is obtained in this range of magnetic Reynolds numbers. Note that this does not preclude a dynamo capability of the flow. Indeed, for comparison, the decay rates for a full numerical MHD simulation of a sphere precessing at an angle of $60^{\circ}$ and a precession rate $\Omega_p=0.3$ \cite[which corresponds to the case studied in][]{Tilgner_2005} are also reported for $E=1/700$. According to \cite{Tilgner_2005}, the dynamo threshold is obtained for $Pm/E \approx 7000$, at least if the role of the small inner core present in his study is negligible. The points represented in figure \ref{cebronfig9}b are close to the points corresponding to elliptical instability flows and do not allow to predict such a dynamo capability. Note that these results are also close to the decay rate reported by \cite{Tilgner_1998} in his study of the kinematic dynamo ability  of the Poincar\'e flow in a precessing spheroid of aspect ratio $c/a=0.9$. Figure \ref{cebronfig9}c is the same as figure \ref{cebronfig9}b except that pseudo-vacuum conditions are used at the outer boundary. Most of the points follow the purely diffusive trend. However, the mode (1,3) seems to leave the purely diffusive behaviour for the larger $Rm$, which is very encouraging. Such a trend leads to expect a dynamo capability of the flow.

\section{Conclusion}

In conclusion, numerical simulations of dynamo are tractable
with the commercial software COMSOL Multiphysics\textsuperscript{\circledR}, based on a finite-element
method. Successfull validations have been obtained for kinematic dynamos on a Ponomarenko-like problem and a Von Karman flow, and also on the thermally driven dynamic dynamo of the \cite{Christensen} benchmark. This finite element approach presents the great advantage of being capable
of dealing easily with complex geometries. In the present study, we
have focused on the MHD flows in a triaxial ellipsoid,
representative of a tidally deformed planetary core. The first MHD
numerical simulations of the elliptical instability under an imposed
magnetic field in this geometry have been presented. Results
regarding the instability growth rate in the presence of Joule
dissipation and the induced magnetic field have been discussed, and
the analytical results coming from local studies have been
confirmed. In the near future, following the parallelization of the
software COMSOL Multiphysics\textsuperscript{\circledR} and the
resulting increased computational power, it is expected that the
elliptical instability will be validated as a dynamo capable forcing
at the planetary scale.

\section{Acknowledgments}
The authors are grateful to C. Nore (LIMSI, Orsay) for fruitful discussions.

\end{document}